\documentstyle[prc,aps,epsfig]{revtex}
\draft
\tighten

\begin{document}

\newcommand{\kpaf}{{{k}^{\prime}_{\alpha}}}
\newcommand{\kpbf}{{{k}^{\prime}_{\beta}}}
\newcommand{\kpgf}{{{k}^{\prime}_{\gamma}}}
\newcommand{\kpao}{{{k^{'0}_{\alpha}}}}
\newcommand{\ka}{{\vec{k}_{\alpha}}}
\newcommand{\kb}{{\vec{k}_{\beta}}}
\newcommand{\kg}{{\vec{k}_{\gamma}}}
\newcommand{\kaf}{{{k}_{\alpha}}}
\newcommand{\kbf}{{{k}_{\beta}}}
\newcommand{\kgf}{{{k}_{\gamma}}}
\newcommand{\kao}{{{k}_{\alpha}^{0}}}
\newcommand{\qpa}{{\vec{q}^{\prime}_{\alpha}}}
\newcommand{\qpaunit}{{\hat{q}^{\prime}_{\alpha}}}
\newcommand{\ppa}{{\vec{p}^{\prime}_{\alpha}}}
\newcommand{\qa}{{\vec{q}_{\alpha}}}
\newcommand{\qaf}{{{q}_{\alpha}}}
\newcommand{\qaunit}{{\hat{q}_{\alpha}}}
\newcommand{\pa}{{\vec{p}_{\alpha}}}
\newcommand{\paf}{{{p}_{\alpha}}}
\newcommand{\Phefour}{P_{\mathrm{^{3}He}}}
\newcommand{\Phfour}{P_{\mathrm{^{3}H}}}
\newcommand{\Ph}{\vec{P}_{\mathrm{^{3}H}}}
\newcommand{\Pheo}{{P}^{0}_{\mathrm{^{3}He}}}
\newcommand{\Pho}{{P}^{0}_{\mathrm{^{3}H}}}
\newcommand{\Phos}{P^{0\;2}_{\mathrm{^{3}H}}}
\newcommand{\Qhepm}{{Q^{H}}}
\newcommand{\Qhepml}{{Q_{H}}}
\newcommand{\Qlepm}{{Q^{L}}}
\newcommand{\Qlepml}{{Q_{L}}}
\newcommand{\statevec}[1]{{\mid{#1}\rangle}}
\newcommand{\phasespace}[1]{{\frac{d^{3}{#1}}{(2\pi)^{3}}}}
\newcommand{\dirac}[1]{{(2\pi)^{3}\delta^{(3)}({#1})}}
\newcommand{\diracfour}[1]{{(2\pi)^{4}\delta^{(4)}({#1})}}
\newcommand{\besselj}[2]{{j_{#1}(#2)}}
\newcommand{\sphar}[3]{{Y_{#1}^{#2}({#3})}}
\newcommand{\cgc}[6]{{\left(\begin{array}{cccc}{#1}&\!\Big{|}&\!
{#3}&\!{#5}\\{#2}&\!\Big{|}&\!{#4}&\!{#6}\end{array}\right)}}
\newcommand{\sixj}[6]{{\left \{ 
\begin{array}{ccc}
{#1}&{#2}&{#3}
\\
{#4}&{#5}&{#6}
\end{array}
\right \}}}
\newcommand{\ninej}[9]{{\left \{ 
\begin{array}{ccc}
{#1}&{#2}&{#3}
\\
{#4}&{#5}&{#6}
\\
{#7}&{#8}&{#9}
\end{array}
\right \}}}
\newcommand{\ip}[2]{{\langle{#1}\mid{#2}\rangle}}
\newcommand{\rme}[3]{{\langle{#1}\| {#2} \| {#3}\rangle}}
\newcommand{\rc}[4]{{[(({#1}){#2}\otimes{#3})_{#4}]}}
\newcommand{\sm}{{\vec{\sigma}}}
\newcommand{\sml}{{\vec{\sigma}_{L}}}
\newcommand{\pol}{{\vec{\epsilon}_{\lambda}}}
\newcommand{\dprod}[2]{{{#1}\cdot {#2}}}
\newcommand{\cprod}[2]{{{#1}\times {#2}}}
\newcommand{\ffone}{{[{\mathbf{1}}]^{0}}}
\newcommand{\fftwo}{{[{\mathbf{\sm}}]^{0,1}}}
\newcommand{\ffthree}{{[{\mathbf{\sm}}]^{2,1}}}
\newcommand{\feynprop}[1]{{S_{F}({#1})}}
\newcommand{\mtri}{{M_{\mbox{\scriptsize t}}}}
\newcommand{\mpp}{{M_{p}}}
\newcommand{\feynsh}[1]{{{#1} \hspace{-1.7mm} \slash}}


\title{\hfill{\small {TRI-PP-01-37}}\\[0.2cm]
Radiative Muon Capture by $\mathrm{^{3}He}$}
\author{Ernest C. Y. Ho}
\address{Department of Physics and Astronomy, University of British 
Columbia, 6224 Agricultural Road, Vancouver, BC, V6T 1Z1, Canada.}

\author{Harold W. Fearing and Wolfgang Schadow \footnote{Current
Address: Cap Gemini Ernst and Young, Hamborner Str. 55, 40472
D\"usseldorf, Germany}}

\address{TRIUMF, 4004 Wesbrook Mall, Vancouver, BC, V6T 2A3, Canada.}

\date{November 29, 2001}

\maketitle

\begin{abstract}
The rate of the nuclear reaction
$\mathrm{^{3}He}+\mu^{-}\rightarrow\mathrm{^{3}H}+\gamma+\nu_{\mu}$
has been calculated using both the elementary particle model (EPM)
approach and the impulse approximation (IA) approach.  Using the EPM
approach, the exclusive statistical radiative muon capture (RMC) rate
for photon energy greater than 57 MeV is found to be
$0.245\;{\mathrm{s}}^{-1}$ and the ordinary muon capture (OMC) rate to
be $1503\;{\mathrm{s}}^{-1}$.  The IA calculation exhibits a slight
dependence on the type of trinucleon wave functions used.  The
difference between the IA and EPM calculation is larger for RMC than
for OMC.  To resolve the difference between the two approaches a more
detailed investigation including meson exchange corrections will be
required.
\end{abstract}


\section{Introduction}

A recent TRIUMF experiment \cite{E592,Wright99} designed to measure
the rate of the radiative muon capture (RMC) reaction
$\mathrm{^{3}He}+\mu^{-}\rightarrow\mathrm{^{3}H}+\gamma+\nu_{\mu}$
has sparked a renewed interest in this process.  Since it is more
sensitive to the nucleon pseudoscalar form factor $g_{P}$ than its
non-radiative counterpart (Refs. \cite{Mukhopadhyay77,Measday01} are
two reviews on ordinary muon capture), it is an ideal candidate for
checking the value of this form factor which is theoretically predicted
by PCAC (Partial Conservation of Axial Current).  With the experiment
on-going, it is necessary to have a modern theoretical calculation of
the process to interpret the anticipated experimental results.  In
this paper, we have calculated the rate of the process using two
perspectives: a nuclear perspective via the elementary particle model
(EPM) and a nucleon perspective via the impulse approximation (IA).
	
The only similar calculation for this process was done by Klieb and
Rood \cite{Klieb81,Kliebthesis} about twenty years ago but the
accuracy of their calculation is constrained by the facts that 1) the
trinucleon wave function they used is inadequate by today's standards,
2) some of the nucleon momentum terms were handled in an approximate
way and 3) they did not use the full Adler and Dothan amplitude but
included only some of the Adler and Dothan terms.  As a consequence of
better computer technology and better methods of calculation of
realistic trinucleon wave functions, it has been possible in this
calculation to improve on the approximations they made. The analogous
nonradiative, or ordinary muon capture (OMC), reaction has been
considered by several authors. See for example
Refs. \cite{Congletonthesis,Congleton93,Congleton96,Govaerts00}, and
references cited therein.

In section \ref{sec:trans_amp} we will briefly discuss various
hypotheses governing the weak hadronic current and the Adler and
Dothan \cite{Adler66} procedure which provides terms in addition to
terms generated by a naive insertion of photons on each particle with
charge or magnetic moment (see also Ref. \cite{Rood65} for early work
on this topic and Ref. \cite{Christillin80} for an alternative
presentation of the Adler and Dothan procedure).  The elementary
particle model approach is discussed in section \ref{sec:epm}, where
both $\mathrm{^{3}He}$ and $\mathrm{^{3}H}$ nuclei are treated as
single entities with internal structure revealed only by the
phenomenological nuclear form factors taken from experiments.  We will
then discuss the impulse approximation approach in section
\ref{sec:ia}.  The essence of the impulse approximation is that it
regards the radiative capture process as taking place on the
constituent nucleons.  Assuming that the nucleons are free, one then
uses a trinucleon wave function to integrate out the internal degrees
of freedom.  The resulting amplitude will be one which only depends on
the EPM, or external, degrees of freedom.  The results will be
presented in section \ref{sec:res} and a summary in section
\ref{sec:summary}.

\section{The transition amplitude}\label{sec:trans_amp}

\begin{figure}[htb]
{
\hspace{40mm}
\psfig{figure=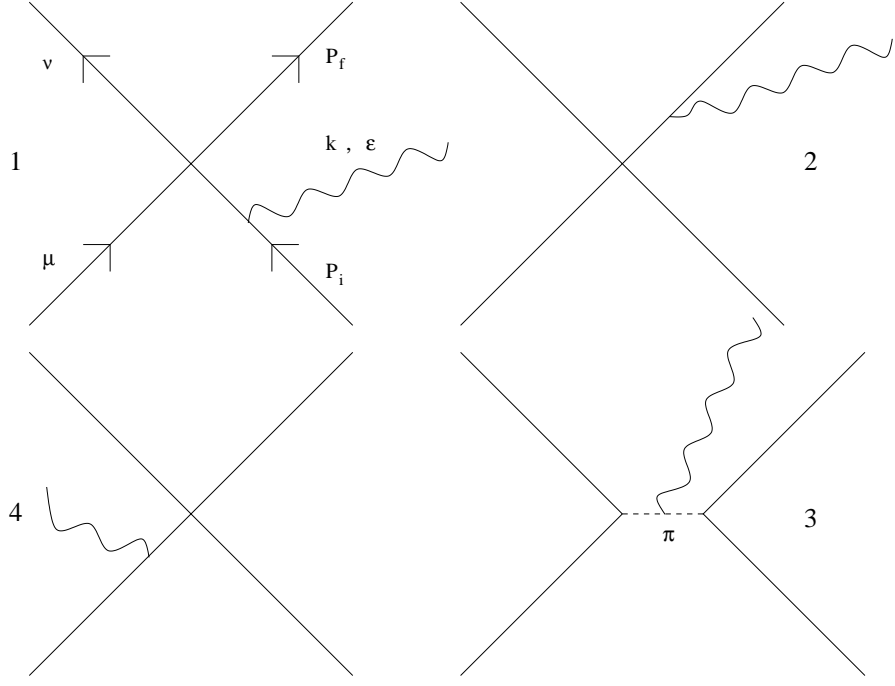,width=90mm,angle=-90}
\vspace{0.75cm}
\caption{The fundamental radiating diagrams\label{rmc_ext_rad}}
}
\end{figure}

The fundamental terms of the transition amplitude are obtained by
inserting a photon on all particles carrying charge or magnetic
moment.  Figure (\ref{rmc_ext_rad}) shows the Feynman diagrams
corresponding to these terms.  Note that this general approach is
common to both EPM and IA so that the form of the transition amplitude
derived here will be applied to the nucleus in the EPM, but to the
nucleon in the IA.

The amplitudes corresponding to the diagrams in Fig. (\ref{rmc_ext_rad}) are
\begin{eqnarray}
M_{1}&=&\bar{u}(\nu)\gamma_{\alpha}(1-\gamma^{5})u(\mu)\bar{u}(P_{f})W^{\alpha}
(\Qhepm)\feynprop{{P}_{i}-{k}}{\mathcal{Q}}_{i}u(P_{i})
\nonumber\\
M_{2}&=&\bar{u}(\nu)\gamma_{\alpha}(1-\gamma^{5})u(\mu)\bar{u}(P_{f})
{\mathcal{Q}}_{f}\feynprop{{P}_{f}+{k}}W^{\alpha}(\Qhepm)u(P_{i})
\nonumber\\
M_{3}&=&\bar{u}(\nu)\gamma_{\alpha}(1-\gamma^{5})u(\mu)\bar{u}(P_{f})\big{\{}
\frac{-i}{m_{\pi}^{2}-(\Qhepm-k)^{2}}\dprod{(-i)(2\Qhepm-k)}{\epsilon}G^{H}_{P}
\big{\}}\frac{\Qhepml^{\alpha}}{m}\gamma^{5}u(P_{i})
\nonumber\\
M_{4}&=&\bar{u}(\nu)\gamma_{\alpha}(1-\gamma^{5})\feynprop{{\mu}-{k}}(-i
\feynsh{\epsilon})u(\mu)\bar{u}(P_{f})W^{\alpha}(\Qlepm)u(P_{i})
\end{eqnarray}
where $S_{F}$ is the Feynman propagator for spin $\frac{1}{2}$
particles
($\feynprop{{P}_{i}-{k}}=\frac{i}{\feynsh{P}_{i}-\feynsh{k}-M_{n}}$
for example), $\nu$, $\mu$ and $k$ are the four-momenta of the
neutrino, muon and the emitted photon respectively and $W^{\alpha}(Q)$
is the weak hadronic vertex which is parameterized by four form factors
\begin{eqnarray}
W^{\alpha}(Q)&=&G_{V}\gamma^{\alpha}+G_{M}i\sigma^{\alpha \beta}
\frac{Q_{\beta}}{2M_{n}}+G_{A}\gamma^{\alpha}\gamma^{5}+G_{P}
\gamma^{5}\frac{Q^{\alpha}}{m}
\\
\sigma^{\alpha \beta}&\equiv&\frac{i}{2}(\gamma^{\alpha}\gamma^{\beta}-
\gamma^{\beta}\gamma^{\alpha})
\end{eqnarray}
with all the $G_{i}$'s functions of $Q^{2}$, the square of the momentum
transfer at the weak hadronic vertex.  Specifically, we denote
$\Qhepm=\mu-\nu$ as the momentum transfer at the hadronic vertex when
one of the hadrons is radiating and $\Qlepm=\mu-\nu-k$  as the momentum
transfer when the lepton is radiating.
${\mathcal{Q}}_{i(f)}=ie_{i(f)}\feynsh{\epsilon}+\frac{\kappa_{i(f)}}{2M_{n}}
\sigma^{\lambda\rho}k_{\rho}\epsilon_{\lambda}$
where $e_{i(f)}, \kappa_{i(f)}$ denote the electric charge and
anomalous magnetic moment of the initial (final) particle.  The
induced pseudoscalar coupling, $G_{P}$, which originates from the pion
pole term has the form \cite{Primakoff79}
\begin{eqnarray}
G_{P}(Q^{2})&=&\frac{2mM_{n}G_{A}(Q^{2})(1+\varepsilon)}{m_{\pi}^{2}-Q^{2}}
\nonumber\\
\varepsilon&=&\frac{m_{\pi}^{2}}{-Q^{2}}\big{\{}1-\frac{G_{\pi}(Q^{2})
/G_{\pi}(0)}{G_{A}(Q^{2})/G_{A}(0)}\big{\}}\label{pionpole}
\end{eqnarray}
which comes from PCAC and the Goldberger-Treiman relation for ordinary
muon capture. The quantity $\varepsilon$ can be regarded as a constant
over the $Q^{2}$ concerned
\cite{Klieb81,Kliebthesis,Congletonthesis,Congleton93}.  The ``PCAC''
value of $G_{P}$ is defined as $\varepsilon=0$.  $G_{\pi}$ is the
pion-i-f coupling constant.  In the EPM, the hadronic vertex is at the
nuclear level.  Therefore, $P_{i}\equiv\Phefour$ (four momentum of the
$\mathrm{^{3}He}$ nucleus) and $P_{f}\equiv\Phfour$ (four momentum of
the $\mathrm{^{3}H}$ nucleus).  In the IA, the hadronic weak
interaction operator acts on the constituent \emph{nucleons} and so
$P_{i}\equiv p$ (four momentum of the proton) and $P_{f}\equiv n$
(four momentum of the neutron).  Note also that $M_{n}$ is the mass of
nucleus in the EPM but the mass of nucleon in the IA and $m$
($m_{\pi}$) is used to denote the mass of muon (charged pion)
throughout.

However, the sum of these four diagrams is not gauge invariant (GI)
and does not satisfy CVC and PCAC by itself.  Extra terms must be
added in order for the whole transition amplitude to be gauge
invariant and to satisfy CVC and PCAC up to a desired order.  The Adler
and Dothan procedure is used to generate these terms up to
${\mathcal{O}}(k^{0})$ via the GI requirement and
${\mathcal{O}}(Q^{0})$ via the CVC and PCAC hypotheses.  The extra
piece of amplitude that is required is
\begin{eqnarray}
\Delta M&=&\Delta M_1+\Delta M_2+\Delta M_3 \label{deltaM}\\
\Delta M_1&=&-\bar{u}(\nu)\gamma_{\alpha}(1-\gamma^{5})u(\mu)\bar{u}
(P_{f})\Big{\{} G_{M}^{L}i\sigma^{\alpha \zeta}\frac{\epsilon_{\zeta}}
{2M_{n}}+G_{P}^{L}\frac{\epsilon_{\alpha}}{m}\gamma^{5}
\Big{\}}u(P_{i}) \label{deltaM1}\\
\Delta M_2&=&\bar{u}(\nu)\gamma_{\alpha}(1-\gamma^{5})u(\mu)\bar{u}(P_{f})
\Big{\{}-2(G_{V}^{\prime}\gamma_{\mu}+G_{A}^{\prime}\gamma_{\mu}\gamma^{5})
(k^{\alpha}\epsilon^{\mu}-k^{\mu}\epsilon^{\alpha})-G_{V}^{\prime}
(\kappa_{f}-\kappa_{i})\frac{2k^{\alpha}}{2M_{n}}i\sigma^{\beta\zeta}
k_{\zeta}\epsilon_{\beta}\Big{\}}u(P_{i}) \label{deltaM2}\\
\Delta M_3&=&\bar{u}(\nu)\gamma_{\alpha}(1-\gamma^{5})u(\mu)\bar{u}(P_{f})
\Big{\{}-2(G_{V}^{\prime}\gamma_{\mu}+G_{A}^{\prime}\gamma_{\mu}\gamma^{5})
g^{\mu \alpha}\dprod{\Qhepm}{\epsilon}
-\frac{2mM_{n}G_{A}^{\prime}(1+\varepsilon)}{m_{\pi}^{2}-\Qlepml^{2}}
\frac{2\dprod{\Qhepm}{\epsilon}}{m}\Qhepml^{\alpha}\gamma^{5} \nonumber \\
&\quad \quad &-G_{M}^{\prime}i\sigma^{\alpha \beta}\frac{\Qhepm_{\beta}}
{2M_{n}} (2\dprod{\Qhepm}{\epsilon})
\Big{\}}u(P_{i}). \label{deltaM3}
\end{eqnarray}
A prime on a form factor denotes the derivative with respect to
$Q^{2}$.  Since all the form factors (except $G_{P}$ which contains a
pole term) are almost linear in $Q^{2}$ over the range of $Q^{2}$
concerned in both the EPM and IA, it does not matter with respect to
which $Q^{2}$ the derivative is taken.  The pion pole terms of $G_{P}$
are treated exactly.

\begin{figure}[htb]
{
\hspace{70mm}
\psfig{figure=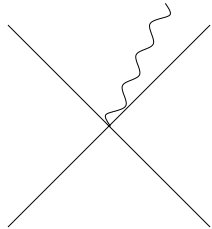,width=30mm,angle=-90}
\vspace{0.75cm}
\caption{The ``minimal coupling'' diagram\label{minimal_coupling_rad}}
}
\end{figure}

The term $\Delta M_1$ in $\Delta M$ arises when one does a minimal
substitution on the hadronic vertex, assuming constant form factors,
and corresponds to the usual fifth diagram included in previous RMC
calculations (Fig.(\ref{minimal_coupling_rad})).  If the nuclei or
nucleons were elementary particles with no form factors, this would be
the only term needed in $\Delta M$ in order to ensure GI, CVC and PCAC
of the amplitude.  The other terms in $\Delta M$ involving derivatives
of form factors are the terms to account for the composite nature of
both the nuclei (in the EPM) and the nucleons (in the IA).  The term
$\Delta M_2$ is demanded by the CVC and PCAC while the terms in
$\Delta M_3$ are demanded by the gauge invariance requirement in the
case when form factors are included \footnote{These are almost all the
terms considered by Adler and Dothan \cite{Adler66} except that we
have not considered the non-Born terms in the pion photoproduction
amplitude which are thought to be small.  Klieb and Rood
\cite{Klieb81,Kliebthesis} did not have the last term in $\Delta M_2$
and the last two terms in $\Delta M_3$.}.

The full amplitude $M \equiv \sum_{i=1-4}M_{i}+\Delta M$ satisfies GI,
CVC and PCAC up to, but not including, terms of ${\mathcal{O}}(kQ)$
and this is the amplitude that we will use for later calculations.

\section{The elementary particle model}\label{sec:epm}

The \emph{elementary particle model} is probably the simplest method
to calculate the RMC rate.  It was first used by Kim and Primakoff
\cite{Kim65,Kim65s} in calculating the beta decay of complex nuclei
and was subsequently used by Fearing \cite{Fearing80} and Klieb and
Rood \cite{Klieb81,Kliebthesis} in their RMC calculations.  In this
approach, both the $\mathrm{^{3}He}$ and $\mathrm{^{3}H}$ are treated
as ``elementary particles'' of spin $\frac{1}{2}$ and isospin
$\frac{1}{2}$ up to a small isospin breaking.  The details of the
structure of the nuclei are encapsulated in the nuclear form factors
which are determined by experiments done on these nuclei.  One of the
major reasons the EPM calculation is easier than the IA is that there
are many fewer degrees of freedom which one has to take care of in the
EPM than in the IA.  The degrees of freedom in the EPM approach are
the four-momenta of $\mathrm{^{3}He}$ ($\Phefour$), $\mathrm{^{3}H}$
($\Phfour$), photon ($k$), neutrino ($\nu$) and muon ($\mu$) together
with their respective spins.  The four-momentum conservation relation
is
\begin{eqnarray}
\Phfour+\nu+k&=&\Phefour+\mu.
\end{eqnarray}
The differential capture rate (photon spectrum) is given by
\begin{eqnarray}
\frac{d\Gamma}{dk}&=&\sum_{\mathrm{photon\;polarization}}\int{
\frac{d(phase\;space)}{dk}\mbox{Tr}(\rho M(\Phfour,\Phefour,\mtri)^
{\dag}M(\Phfour,\Phefour,\mtri))}\label{diff_cap_rate}
\end{eqnarray}
where $d(phase\;space)$ is the differential phase space factor
\begin{eqnarray}
d(phase\;space)&=&C|\phi_{\mu}(0)|^{2}\frac{|
\mathrm{V_{ud}|^{2}G_{F}^{2}}}{2}\phasespace{\Ph}\phasespace{
\vec{\nu}}\frac{1}{k}\phasespace{\vec{k}}\diracfour{\mu+\Phefour-
\Phfour-k-\nu}
\nonumber\\
&=&C|\phi_{\mu}(0)|^{2}\frac{1}{2\pi^{5}}\frac{|
\mathrm{V_{ud}|^{2}G_{F}^{2}}}{2}\frac{2\Pho}{k}|2k(1-
\cos(\theta))-2(m+\mtri)|^{-1}\nu^{2}(k,\theta)d\hat{\nu}k^{2}dkd\hat{k}
\\
\cos(\theta)&=&\dprod{\hat{\nu}}{\hat{k}}
\\
\nu(k,\theta)&=&\frac{2k(m+\mtri)-m^{2}-2m\mtri}{2k(1-\cos(\theta))-
2(m+\mtri)}
\\
\mtri&=&2808.66\;\mathrm{MeV}.
\end{eqnarray}
Here $\phi_{\mu}(0)$ denotes the muon wave function at the origin,
$C=0.9788$ \cite{Congletonthesis,Congleton93} is the correction factor
that accounts for the non-pointlike nature of the nucleus, and
$\mathrm{V_{ud}}$ is the CKM matrix element which connects the up and
down quark, with $\mathrm{V_{ud}}=0.9735\pm0.0008\cite{Groom00}$.
$\mathrm{G_{F}}$ is the Fermi coupling constant, $\rho$ is the density
matrix which describes the initial spin configuration of the muonic
atom, and $M(\Phfour,\Phefour,\mtri)$ is the transition amplitude $M$
that was discussed in section (\ref{sec:trans_amp}) with the following
name changes
\begin{eqnarray}
u(P_{i})&\rightarrow&u(\Phefour)
\nonumber\\
\bar{u}(P_{f})&\rightarrow&\bar{u}(\Phfour)
\nonumber\\
M_{n}&\rightarrow&\mtri\;\mathrm{on\;the\;hadronic\;vertex}
\nonumber\\
G_{i}&\rightarrow&F_{i}
\end{eqnarray}
where $F_{i}, i=V,M,A$ are the nuclear form factors which are
parameterized as
\begin{eqnarray}
F_{i}&=&F_{i}(0)(1+\frac{1}{6}R_{i}^{2}Q^{2})\label{nuclear_ff}
\end{eqnarray}
with 
$F_{V}(0)=1,\;
R_{V}=1.94\;{\mathrm{fm}}\cite{Congletonthesis},\;
F_{M}(0)=\kappa_{{\mathrm{^{3}He}}}-\kappa_{{\mathrm{^{3}H}}}
=-8.369\;{\mathrm{n.m}}-7.913\;{\mathrm{n.m}},\;
R_{M}=1.72\;{\mathrm{fm}}\cite{Congletonthesis},\;
F_{A}(0)=1.212\pm0.004\;\cite{Congleton93},\;
R_{A}=1.703\;{\mathrm{fm}} \cite{Congletonthesis}$.

The photon polarization vector $\pol$, which is defined by
$\pol\equiv\frac{1}{\sqrt{2}}(\hat{x}+i\lambda\hat{y})$, with
$\hat{k}=\hat{z}$, has the property
\begin{eqnarray}
\cprod{\vec{k}}{\pol}=-i\lambda k \pol.
\end{eqnarray}
Thus the $\sum_{\mathrm{photon\;polarization}}$ in Eq.
(\ref{diff_cap_rate}) can be replaced by $\sum_{\lambda=-1,+1}$.

The big advantage of the EPM is its simplicity. It also includes, to
some extent at least, part of the meson exchange corrections which are
missing in the IA. It however does have a major flaw as applied to a
two step process such as RMC in that the intermediate states have to
be treated as elementary particles as well. Thus effects coming from
excitation of the intermediate nucleus, which are implicitly partially
included in the IA, are not included in the EPM. A full investigation
of this is beyond the scope of this paper, though see
Ref. \cite{Klieb85}. However one should keep this in mind as a caveat
with respect to the EPM.

\section{The impulse approximation}\label{sec:ia}

The impulse approximation method provides a simple ``microscopic''
picture of the nuclear reaction in terms of nucleons.  In this
picture, the constituent nucleons inside the nucleus are approximated
as free (this is probably a good approximation as the binding energy
of the trinucleon system is $\sim$8 MeV, which is about 0.3$\%$ of the
mass of the nucleus) and the nuclear reaction
$\mathrm{^{3}He}+\mu^{-}\rightarrow\mathrm{^{3}H}+\nu_{\mu}+\gamma$ is
viewed as the sum of its nucleon counterparts $p+\mu^{-}\rightarrow
n+\nu_{\mu}+\gamma$ (that is, only one-body currents are considered
and two-body meson exchange currents are neglected or put in later as
a correction).  The extra degrees of freedom which arise from
considering each nucleon of the nucleus instead of treating the
nucleus as a whole are integrated out using realistic tri-nucleon
wave functions.  Given these assumptions, the whole problem boils down
to separating the EPM and non-EPM degrees of freedom and finding the
IA equivalent (denote it as $M_{ia}$) of $M(\Phfour,\Phefour,\mtri)$
in the EPM.  More explicitly, we want the $M(\Phfour,\Phefour,\mtri)$
in Eq. (\ref{diff_cap_rate}) to be replaced with $M_{ia}$ in
order to find the IA version of $\frac{d\Gamma}{dk}$.  The
relationship between $M(\Phfour, \Phefour, \mtri)$ and $M_{ia}$ is:
\begin{eqnarray}
M(\Phfour,\Phefour,\mtri)&\leftrightarrow&{\mathbf{3}}\int\big{\{}
\dirac{\pa^{\prime}-\pa}\dirac{\qpa-\qa+\frac{2}{3}(\vec{\nu}+\vec{k}-
\vec{\mu})}\psi^{*}_{\mathrm{^{3}H}}(\pa^{\prime},\qpa)\psi_{
\mathrm{^{3}He}}(\pa,\qa)\times
\nonumber\\
&&M(\kpaf,\kaf,\mpp)\phasespace{\qpa}\phasespace{\pa^{\prime}}
\phasespace{\qa}\phasespace{\pa}\big{\}}\equiv M_{ia}
\label{iatoepm}
\end{eqnarray}
where $\kaf, \kbf, \kgf$ ($\kpaf, \kpbf, \kpgf$) denote the four
momenta of the three initial (final) nucleons $\alpha$, $\beta$ and
$\gamma$.  The (three) momentum transformation separating the EPM and
non-EPM degrees of freedom is
\begin{eqnarray}
\vec{P}&=&\ka+\kb+\kg
\nonumber\\
\qa&=&\frac{2}{3}\ka-\frac{1}{3}\kb-\frac{1}{3}\kg
\nonumber\\
\pa&=&\frac{1}{2}\kb-\frac{1}{2}\kg
\end{eqnarray}
for the initial nucleus, with an identical transformation law for the
final nucleus.  It is obvious that $\vec{P}$ (a EPM degree of freedom)
is the center of mass momentum vector of the nucleus and $\qa$ (a
non-EPM degree of freedom) is the momentum of nucleon $\alpha$
(spectator) with respect to the center of mass momentum of the other
two nucleons (subsystem) while $\pa$ (a non-EPM degree of freedom) is
the momentum of particle $\beta$ with respect to particle
$\gamma$\footnote{$\pa$ and $\qa$ will sometimes be denoted as
$\vec{p}$ and $\vec{q}$ when no confusion arises.}.  The $\mathbf{3}$
comes from the antisymmetrization of the wave function and it allows
one to let the current operator act on a particular nucleon (chosen to
be nucleon $\alpha$) three times instead of acting on each nucleon of
the nucleus.

The parameterization of the nucleon form factors is exactly the same as
that of the nuclear ones (see Eq. (\ref{nuclear_ff})).  For
convenience in notation, we change $F_{i}\rightarrow g_{i}$ and
$R_{i}\rightarrow r_{i}$ in Eq. (\ref{nuclear_ff}) to denote the
nucleon case. The various parameters for nucleon form factors are
$g_{V}(0)=1, 
r_{V}=0.7589\;{\mathrm{fm}}\;\cite{Congletonthesis},\;g_{M}(0)=
\kappa_{p}-\kappa_{n},\;r_{M}=0.8781\;{\mathrm{fm}}\;\cite{Congletonthesis},
\;g_{A}(0)=-1.267\pm0.0035\;\cite{Groom00},
\;r_{A}=0.6580\;{\mathrm{fm}}\;\cite{Congletonthesis}$.

The momentum space trinucleon wave functions
\cite{Schadow00,Schadowthesis} are realistic
wave functions derived from the Faddeev equation (see, for example,
Ref. \cite{Glockle}) with different model potentials. Each one of them
has 22 channels which contain all possible states up to and including
$J=2$, where $J$ is the total angular momentum of the subsystem
particles.  They can be written as,
\begin{eqnarray}
\statevec{\Psi}&=&\sum_{i_{c}}\psi_{i_c}(\paf,\qaf)\statevec{i_{c}}
\statevec{\vec{P}}\label{wavefunc}
\end{eqnarray} 
where $i_{c}$ is the channel number and $\statevec{\vec{P}}$ is the
center of mass momentum of the trinucleon system which we will not
write explicitly from now on. The coupling scheme of the channels is,
\begin{eqnarray}
\statevec{i_{c}}&=&\statevec{((L_{\alpha}\,l_{\alpha}){\mathcal{L}}_
{\alpha},(S_{\alpha} s_{\alpha}){\mathcal{S}}_{\alpha}){\mathcal{J}}}
\statevec{(I_{\alpha}i_{\alpha}){\mathcal{I}}}
\nonumber\\
&\equiv&\statevec{i_{c}({\mathcal{J}})}\statevec{i_{c}({\mathcal{I}})}.
\end{eqnarray}
The spin and angular momentum part is given by
$\statevec{((L_{\alpha}\,l_{\alpha}){\mathcal{L}}_{\alpha},(S_{\alpha}
s_{\alpha}){\mathcal{S}}_{\alpha}){\mathcal{J}}}$ with $L_{\alpha}$
($S_{\alpha}$) the angular momentum (spin) of subsystem $(\beta
\gamma)$ and $l_{\alpha}$ ($s_{\alpha}$) the angular momentum (spin)
of particle $\alpha$ (the spectator particle). These are coupled to
form ${\mathcal{L}}_{\alpha}$ (${\mathcal{S}}_{\alpha}$) and then to
${\mathcal{J}}=\frac{1}{2}$.  The
$\statevec{(I_{\alpha}i_{\alpha}){\mathcal{I}}}$ is the isospin part,
with $I_{\alpha}$ being the subsystem isospin which couples with
$i_{\alpha}$, the spectator isospin, to form
${\mathcal{I}}=\frac{1}{2}$.

It is now clear that $\psi_{\mathrm{^{3}He}}(\pa,\qa)$ in Eq.
(\ref{iatoepm}) is just the $\langle\hat{p},\hat{q}|$ projection
of Eq. (\ref{wavefunc})
\begin{eqnarray}
\psi_{\mathrm{^{3}He}}(\pa,\qa)&=&\sum_{i_{c}}\psi_{i_{c}}
(\paf,\qaf)\ip{\hat{p},\hat{q}}{i_{c}}
\end{eqnarray}
where
\begin{eqnarray} 
\ip{\hat{p},\hat{q}}{i_{c}}&=&
\Big{(}\sphar{L\,l}{\mathcal{L}}{\hat{p},\hat{q}}\otimes\chi_{Ss}^{
\mathcal{S}}\Big{)}_{\mathcal{J}}^{M_{\mathcal{J}}}\eta_{Ii}^{{
\mathcal{I}}M_{{\mathcal{I}}}}.
\end{eqnarray}
Note that Eq. (\ref{bipolarhar}) below defines the bipolar harmonic
$\sphar{L\,l}{\mathcal{L}}{\hat{p},\hat{q}}$.  $\chi$ and $\eta$ in
the above equation are spinors and isospinors respectively. Their
coupling method is exactly the same as the bipolar harmonic in
Eq. (\ref{bipolarhar}).

To put $M_{ia}$ in Eq. (\ref{iatoepm}) into a useful format, we
have to expand $M(\kpaf,\kaf,\mpp)$ non-relativistically in powers of
the struck nucleon momentum $\ka$ (which equals $\qa$ upon setting the
initial center of mass momentum of the trinucleon zero) and the
$\delta$ functions into angular momentum eigenstates.  Upon setting
$\vec{\mu}=0$ and denoting $\vec{s}\equiv\vec{\nu}+\vec{k}$ the
$\delta$ functions can be expanded as (see Ref. \cite{Arfken} for
definitions of spherical Bessel functions $\besselj{l}{x}$),
\begin{eqnarray}
\dirac{\ppa-\pa}&=&(2\pi)^{3}\frac{\delta(p_{\alpha}^{\prime}-
p_{\alpha})}{p_{\alpha}^{2}} \sum_{l}(-1)^{l} \sqrt{2l+1}
\sphar{l\,l}{0\,0}{\hat{p_{\alpha}^{\prime}},\hat{p_{\alpha}}}
\\
\dirac{\qpa-\qa+\frac{2}{3}\vec{s}}&=&
\sum_{\stackrel{l_{i},m_{3}}{i=1-3}}
\sqrt{\frac{(4\pi)^{5}(2l_{1}+1)(2l_{2}+1)}{(2l_{3}+1)}}
\cgc{l_{3}}{0}{l_{1}}{0}{l_{2}}{0}
i^{l_{1}-l_{2}+l_{3}}
\sphar{l_{1}\,l_{2}}{l_{3}\,-m_{3}}{\qpaunit,\qaunit}
\sphar{l_{3}}{-m_{3}}{\hat{s}}^{*}
\nonumber\\
&&\times\int\besselj{l_{1}}{q^{\prime}_{\alpha} r}
\besselj{l_{2}}{q_{\alpha} r} 
\besselj{l_{3}}{\frac{2}{3}sr}r^{2}dr 
\\
\sphar{l_{1}\,l_{2}}{l_{3}\,m_{3}}{\hat{x},\hat{y}}&\equiv&
\sum_{m_{1},m_{2}}\cgc{l_{3}}{m_{3}}{l_{1}}{m_{1}}{l_{2}}{m_{2}}
\sphar{l_{1}}{m_{1}}{\hat{x}}\sphar{l_{2}}{m_{2}}{\hat{y}}\label{bipolarhar}
\end{eqnarray}
The notations used here are the same as Brink and Satchler
\cite{Brink} except that the Clebsch-Gordon coefficients are denoted
by $\cgc{J}{M}{J_{1}}{M_{1}}{J_{2}}{M_{2}}$ as opposed to ${\ip{J
M}{J_{1} J_{2} M_{1} M_{2}}}$.  The next thing after the expansion is
to couple all the spin and angular momentum operators in Eq.
(\ref{iatoepm}) into tensors of rank 0 or 1.  Since the total angular
momentum of both initial and final states is $\frac{1}{2}$, there is
no need to couple the operators into tensor of other ranks.  There is
also no need to couple operators into odd parity quantities as both
the initial and final states are of even parity.  Recognizing the
total angular momentum of the trinucleon system in the IA as the spin
in the EPM, one can easily see that any operators of rank 1 in the IA
correspond to (within a factor) $\sm$ matrices in the EPM and
operators of rank 0 in the IA correspond to identity hadronic
operators in the EPM.

All the coefficients in $M(\kpaf,\kaf,\mpp)$ are expanded to
${\mathcal{O}}(\frac{\vec{q}}{\mpp})$ except that of $g_{P}$. The
kinematic endpoints of RMC are quite close to the poles of $g_{P}$ and
thus might make its value large at those places.  Therefore,
coefficients of $g_{P}$ are expanded to
${\mathcal{O}}((\frac{\vec{q}}{\mpp})^{2})$.

The correspondence between the IA and the EPM for all forms of
operators up to first order in momentum is shown below.  Operators of
higher order in momentum will not be shown owing to the lack of
space.  Note that the $[\ldots]$'s are actually reduced matrix
elements between the initial and final states (i.e. results of
integration of ``internal'' degrees of freedom) and the numbers inside
denote some specific spin and angular momentum combination.  They will
be defined in Eq. (\ref{besseltran}).  For now, it is sufficient
to note that the first digit of $[\ldots]$ is related to the nucleon
momentum $\vec{q}$ (for example 0$\sim$ no nucleon momentum, 1$\sim
\vec{q}$) and the second digit comes from the spherical harmonics of
the $\delta$ function.  These two terms couple together to an angular
momentum value represented by the third digit.  The fourth digit is
related to the hadronic spin matrix and the subscript is the rank of
the whole reduced matrix element.  $\mathbf{1}$ (or sometimes denoted
as $\mathbf{1}_{0}$) is defined as the hadronic identity matrix
element in both the IA and the EPM. It differs from $\ffone$ (see
Ref. \cite{Delorme}), although they are related.
\begin{eqnarray}
\mathrm{IA}&&\mathrm{after\;coupling\;and\;reexpressing\;in\;EPM\;format}
\nonumber\\
\mathbf{1}&\leftrightarrow& 
\rc{0,0}{0}{0}{0}\mathbf{1}
\\
\dprod{\sm}{\vec{v}}&\leftrightarrow&
-\frac{3}{\sqrt{2}}\rc{0,2}{2}{1}{1}\dprod{\sm}{\hat{s}}\dprod{\vec{v}}
{\hat{s}}+\{\frac{1}{\sqrt{2}}\rc{0,2}{2}{1}{1}+\rc{0,0}{0}{1}{1}\}
\dprod{\sm}{\vec{v}}\label{iafirst}
\\
\dprod{\sm}{\vec{q}}&\leftrightarrow&
\{-\sqrt{\frac{5}{3}}\rc{1,1}{2}{1}{1}-\frac{1}{2}\rc{1,1}{1}{1}{1}-
\frac{1}{\sqrt{3}}\rc{1,1}{0}{1}{1}\}\dprod{\sm}{\hat{s}}
\\
\dprod{\vec{q}}{\vec{v}}&\leftrightarrow&-\frac{1}{\sqrt{3}}
\rc{1,1}{0}{0}{0}\dprod{\vec{v}}{\hat{s}}-
\frac{i}{\sqrt{2}}\rc{1,1}{1}{0}{1}\dprod{\sm}{\cprod{\vec{v}}{\hat{s}}}
\\
\dprod{\cprod{\sm}{\vec{q}}}{\vec{v}}&\leftrightarrow&\{-\sqrt{\frac{5}{12}}
\rc{1,1}{2}{1}{1}+\frac{1}{2}\rc{1,1}{1}{1}{1}+
\sqrt{\frac{1}{3}}\rc{1,1}{0}{1}{1}\}\dprod{\cprod{\vec{v}}{\hat{s}}}{\sm}+
\nonumber\\
&&i\sqrt{\frac{2}{3}}\rc{1,1}{1}{1}{0}\dprod{\vec{v}}{\hat{s}}
\\
\dprod{\vec{q}}{\vec{v}}\dprod{\sm}{\vec{u}}&\leftrightarrow&
\{\sqrt{\frac{1}{15}}\rc{1,1}{2}{1}{1}-
\sqrt{\frac{1}{3}}\rc{1,1}{0}{1}{1}-
\frac{1}{2}\sqrt{\frac{2}{5}}\rc{1,3}{2}{1}{1}\}\dprod{\sm}
{\vec{u}}\dprod{\vec{v}}{\hat{s}}+
\nonumber\\
&&\{-\sqrt{\frac{3}{20}}\rc{1,1}{2}{1}{1}+\frac{1}{2}\rc{1,1}{1}{1}{1}-
\frac{1}{2}\sqrt{\frac{2}{5}}\rc{1,3}{2}{1}{1}\}\dprod{\sm}
{\vec{v}}\dprod{\vec{u}}{\hat{s}}+
\nonumber\\
&&\{-\sqrt{\frac{3}{20}}\rc{1,1}{2}{1}{1}-\frac{1}{2}\rc{1,1}{1}{1}{1}-
\frac{1}{2}\sqrt{\frac{2}{5}}\rc{1,3}{2}{1}{1}\}\dprod{\sm}{\hat{s}}
\dprod{\vec{u}}{\vec{v}}+
\nonumber\\
&&\sqrt{\frac{5}{2}}\rc{1,3}{2}{1}{1}\dprod{\sm}{\hat{s}}\dprod{\vec{u}}
{\hat{s}}\dprod{\vec{v}}{\hat{s}}+i\sqrt{\frac{1}{6}}\rc{1,1}{1}{1}{0}
\dprod{\vec{u}}{\cprod{\vec{v}}{\hat{s}}}\label{ialast}
\end{eqnarray} 
Note that while $\sm$ on the left hand side acts on the spin of the
spectator nucleon, $\sm$ on the right acts on the entire
\emph{nucleus} and that $\vec{u}$ and $\vec{v}$ are mutually commuting
vectors that are \emph{not} concerned with the internal momenta
(i.e. not $\vec{p}$ nor $\vec{q}$) and commute with $\sm$.  Using
Delorme's \cite{Delorme} notation, $\ffone = \rc{0,0}{0}{0}{0}$,
$\fftwo=\rc{0,0}{0}{1}{1}$, $\ffthree=-\rc{0,2}{2}{1}{1}$,
$[\sm]^{+}=\fftwo+\sqrt{2}\ffthree$ and
$[\sm]^{-}=\fftwo-\frac{1}{\sqrt{2}}\ffthree$.  The precise
relationship between $[i\vec{P}]^{1,1}$ and the reduced matrix
elements defined here is unclear but it has a magnitude of the order
of $\rc{1,1}{1}{0}{1}$ or $\rc{1,1}{1}{1}{1}$.  As one will see later,
these two matrix elements are very small.

The definition of $\rc{a,b}{c[\bar{a}]}{d}{e}$ (a function of 
$s\equiv\|\vec{\nu}+\vec{k}\|$) is
\begin{eqnarray}
\nonumber
\rc{a,b}{c[\bar{a}]}{d}{e}&\equiv&
\end{eqnarray}
\begin{eqnarray}
&&{\mathbf{3}}\frac{1}{2\pi^{5}}\rme{\frac{1}{2}}{T_{e}}{\frac{1}{2}}^{-1}
\rme{\frac{1}{2}}{\vec{\tau}}{\frac{1}{2}}^{-1}\sum_{i_{c},i_{c}^
{\prime}}\sum_{l_{1},l_{2},L_{1}}(-1)^{L_{1}}i^{\;l_{1}-l_{2}+b}
\int\Big{\{} p^{2}dp\: r^{2}dr \besselj{b}{\frac{2}{3}sr}
\{\psi^{*}_{i_{c}^{\prime}}(p,q^{\prime})
\besselj{l^{\prime}}{q^{\prime}r} q^{\prime 2}dq^{\prime}\}
\nonumber\\
&&\{\psi_{i_{c}}(p,q) \besselj{l_{2}}{qr}
q^{2+\bar{a}}dq\}\Big{\}}
\cgc{b}{0}{l_{1}}{0}{l_{2}}{0}F(l_{1},l_{2}; a, b, c; i_{c}^{\prime}, i_{c})
\sqrt{\frac{(2L_{1}+1)(2l_{2}+1)(2l_{1}+1)}{2b+1}}
\nonumber\\
&&\rme{i_{c}^{\prime}({\mathcal{J}})}
{\{ (\sphar{L_{1},L_{1}}{0}{\hat{p}^{\prime},\hat{p}}\otimes
\sphar{l^{\prime},l}{c}{\hat{q}^{\prime},\hat{q}})\otimes({
\mathbf{1}}_{0}\otimes T_{d})\}_{e}} {i_{c}({\mathcal{J}})}
\rme{i_{c}^{\prime}({\mathcal{I}})}{({\mathbf{1}}_{0}\otimes
\vec{\tau})}{i_{c}({\mathcal{I}})}\label{besseltran}
\end{eqnarray}
where $T_{d(e)}=\mathbf{1}_{0}$ for $d(e)=0$ and $\sm$ for $d(e)=1$;
$\vec{\tau}$ is the isospin operator.  Notice $\bar{a}$ specifies the
(mass) dimension of the matrix element.  When $\bar{a}$ is not shown
explicitly on a reduced matrix element, $\bar{a}=a$; that is
$\rc{a,b}{c}{d}{e}\equiv\rc{a,b}{c[a]}{d}{e}$.  $F(l_{1},l_{2}; a, b,
c; i_{c}^{\prime}, i_{c})$ is defined as
\begin{eqnarray}
F(l_{1},l_{2}; a, b, c; i_{c}^{\prime}, i_{c})&\equiv&
\cgc{c}{m_{c}}{a}{m_{a}}{b}{m_{b}}^{-1}
\int
\sphar{l^{\prime},l}{c,m_{c}}{\hat{\upsilon},\hat{\chi}}^{*}
\sphar{0,a}{a,m_{a}}{\hat{\upsilon},\hat{\chi}}
\sphar{l_{1},l_{2}}{b,m_{b}}{\hat{\upsilon},\hat{\chi}}d{\hat{
\upsilon}}d{\hat{\chi}}
\nonumber\\
&=&\frac{1}{4\pi}(-1)^{l_{1}+l_{2}+b}\sixj{l_{1}}{l_{2}}{b}{a}{c}{l}
\cgc{l}{0}{l_{2}}{0}{a}{0}
\sqrt{\frac{(2l_{2}+1)(2a+1)}{2b+1}}
\delta_{{l_{1}}{l^{\prime}}}
\end{eqnarray} 
$\hat{\upsilon}, \hat{\chi}$ being some dummy angular variables.  

$\rme{i_{c}^{\prime}({\mathcal{J}})}
{\{ (\sphar{L_{1},L_{1}}{0}{\hat{p}^{\prime},\hat{p}}\otimes
\sphar{l^{\prime},l}{c}{\hat{q}^{\prime},\hat{q}})\otimes({\mathbf{1}}_{0}
\otimes T_{d}) \}_{e}} {i_{c}({\mathcal{J}})}$ is the spin and angular 
momentum part of the reduced matrix element between the helion and 
triton channels. Its calculation is tedious but standard.
\begin{eqnarray}
\nonumber
\rme{i_{c}^{\prime}({\mathcal{J}})}
{\{ (\sphar{L_{1},L_{1}}{0}{\hat{p}^{\prime},\hat{p}}\otimes
\sphar{l^{\prime},l}{c}{\hat{q}^{\prime},\hat{q}})\otimes({
\mathbf{1}}_{0}\otimes T_{d}) \}_{e}}
{i_{c}({\mathcal{J}})}&=&
\end{eqnarray}
\begin{eqnarray}
&&4\pi(-1)^{{\mathcal{L^{\prime}}}+2l+c+L^{\prime}+{\mathcal{S^{\prime}}}+
\frac{1}{2}+S^{\prime}+d}\big{\{}(2{\mathcal{L}}+1)(2{\mathcal{L^{\prime}}}+
1)(2{\mathcal{S}}+1)(2{\mathcal{S^{\prime}}}+1)
(2c+1)(2e+1)\big{\}}^{\frac{1}{2}}
\nonumber\\
&&\ninej{\frac{1}{2}}{\frac{1}{2}}{e}{{\mathcal{L^{\prime}}}}
{{\mathcal{L}}}{c}{{\mathcal{S^{\prime}}}}{{\mathcal{S}}}{d}
\sixj{{\mathcal{L}}}{{\mathcal{L^{\prime}}}}{c}{l^{\prime}}{l}{L^{\prime}}
\sixj{{\mathcal{S}}}{{\mathcal{S^{\prime}}}}{d}{\frac{1}{2}}{\frac{1}{2}}
{S^{\prime}}\delta_{S S^{\prime}}
\rme{\frac{1}{2}}{T_{d}}{\frac{1}{2}}\rme{L^{\prime}\; 0}
{\sphar{L_{1},L_{1}}{0}{\hat{p}^{\prime},\hat{p}}}{0\; L}.
\end{eqnarray}
Note also that
\begin{eqnarray}
\sum_{L_{1}}(-1)^{L_{1}}\sqrt{2L_{1}+1}
\rme{L^{\prime}\;0}{\sphar{L_{1}, L_{1}}{0}{\hat{p}^{\prime},\hat{p}}}
{0\;L}&=&\frac{1}{4\pi}\delta_{L L^{\prime}}.
\end{eqnarray}
$\rme{i_{c}^{\prime}({\mathcal{I}})}{\mathbf{1}_{0}\otimes\vec{\tau}}
{i_{c}({\mathcal{I}})}$ is the isospin contribution of the reduced matrix 
element, which equals 
\begin{eqnarray}
2(-1)^{I^{\prime}}\sixj{1/2}{1/2}{1/2}{1/2}{1/2}{I^{\prime}}
\delta_{I I^{\prime}}\rme{\frac{1}{2}}{\vec{\tau}}{\frac{1}{2}}.
\end{eqnarray}
By following through all the procedures mentioned in this section, it
is possible to match the IA amplitude piece by piece with its EPM
counterparts and thus make a direct comparison between each piece.  In
other words, we have arranged the non-zero terms of $M_{ia}$ to depend
only on the EPM degrees of freedom.  We then obtain
$\frac{d\Gamma}{dk}$ via Eq. (\ref{diff_cap_rate}) with
$M(\Phfour,\Phefour,\mtri)$ replaced by $M_{ia}$.

\section{Results and Discussion}\label{sec:res}

\subsection{Results}\label{subsec:res}

\begin{figure}[htb]
{
\hspace{20mm}
\psfig{figure=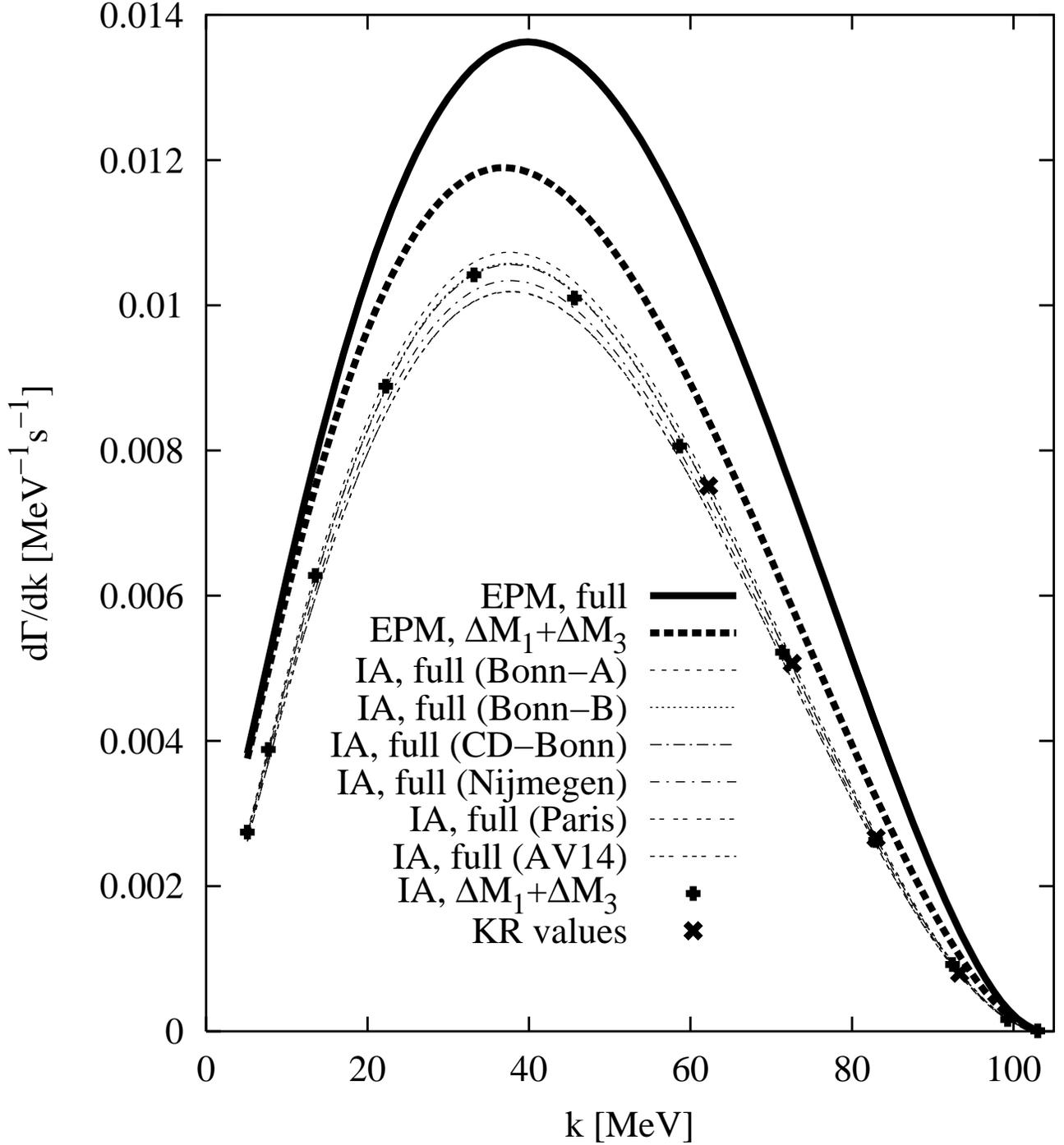,width=190mm,angle=-90}
\vspace{0.75cm}
\caption{ 
RMC photon spectra from two EPM calculations, one with the
full Adler and Dothan amplitude $\Delta M$ and the other with only the
terms $\Delta M_1+\Delta M_3$ necessary for gauge invariance.  Also
included are RMC photon spectra from the full IA calculation using
various model potentials, plus one example with only $\Delta
M_1+\Delta M_3$.  All wave functions used have 22 Faddeev components
and the permutation is projected on the same set of states. The
infrared divergent part is not shown.  The KR values come from Klieb
and Rood \protect\cite{Klieb81,Kliebthesis} and are their IA
calculation results. \label{rmcfig}} 
}
\end{figure}

We have calculated the rate for OMC and the photon spectrum and
integrated rate for RMC in both IA and EPM approaches, using the
formalism described in the preceding sections. Table (\ref{omctable})
shows various RMC and OMC statistical rates, with the final results
given in the last three columns. Figure (\ref{rmcfig}) shows the IA
RMC spectra for various wave functions together with the EPM spectrum.

One sees immediately that the IA results are significantly lower than
the EPM results both for OMC and RMC. This is consistent with the OMC
results of Refs. \cite{Congletonthesis,Congleton93}. For OMC the
difference is understood to arise from various meson exchange
corrections which are included implicitly in the EPM approach but
which must be put into the IA by hand as specific
corrections. \cite{Congleton96}. Presumably a similar explanation
holds for RMC.

\begin{table}[htb]
\hspace{5mm}
\begin{tabular}[c]{|c|c|c|c|c|c|c|}
\hline
Potentials &
\multicolumn{4}{c|}{OMC rate (statistical) ($\mathrm{s^{-1}}$)} &
\multicolumn{2}{c|}{$ \Gamma_{stat}^{rmc}\;(\mathrm{s^{-1}})$}    \\ \hline
 & ${\mathcal{O}}(\frac{\vec{q}}{\mpp})$; $\frac{\vec{\nu}}{3}$
 & ${\mathcal{O}}(\frac{\vec{q}}{\mpp})$; full &  ${\mathcal{O}}(
\frac{\vec{q}^{2}}{\mpp^{2}})$; $\frac{\vec{\nu}}{3}$ &${\mathcal{O}}(
\frac{\vec{q}^{2}}{\mpp^{2}})$; full & $k>$ 5MeV & $k>$ 57MeV    \\ \hline
Bonn-A  &  $\;\;1368.6$ & $1368.1$ 
	&  $1367.8$     & $1357.9$ & $0.6255$ & $0.1691$     \\ \hline
Bonn-B  &  $\;\;1341.3$ & $1340.8$  
	&  $1340.4$     & $1330.7$ & $0.6164$ & $0.1666$    \\ \hline
CD-Bonn &  $\;\;1336.1$ & $1335.7$  
	&  $1335.3$     & $1326.1$ & $0.6153$ & $0.1663$    \\ \hline
Nijmegen I&  $\;\;1298.0$ & $1297.5$  
	&  $1297.1$     & $1288.2$ & $0.6023$ & $0.1626$    \\ \hline
Paris   &  $\;\;1270.9$ & $1270.4$  
	&  $1270.1$     & $1260.3$ & $0.5932$ & $0.1602$    \\ \hline
AV14    &  $\;\;1271.0$ & $1270.6$  
	&  $1270.2$     & $1260.0$ & $0.5929$ & $0.1601$    \\ \hline\hline
EPM     &  \multicolumn{4}{c|}{$1503$}    & $0.8263$&  $0.2451$ \\ \hline 
\end{tabular}
\vspace{0.5cm}
\caption{\label{omctable}Various OMC and RMC statistical rates
calculated with $g_{P}$ or $F_{P}$ at the PCAC value.  The numbers in
the second column are the values obtained by using the
``$\frac{\vec{\nu}}{3}$'' prescription up to
${\mathcal{O}}(\frac{\vec{q}}{\mpp})$ terms.  The numbers in the third
column are obtained via the correct approach up to
${\mathcal{O}}(\frac{\vec{q}}{\mpp})$ terms.  The fourth column has
values of the OMC rates using ``$\frac{\vec{\nu}}{3}$'' prescription
up to ${\mathcal{O}}((\frac{\vec{q}}{\mpp})^{2})$ terms.  Numbers in
the fifth column are the values obtained by the correct approach up to
${\mathcal{O}}((\frac{\vec{q}}{\mpp})^{2})$ terms; the sixth and
rightmost columns contain $\Gamma_{stat}^{rmc}(k>5\;\mathrm{MeV})$ and
$\Gamma_{stat}^{rmc}(k>57\;\mathrm{MeV})$ respectively.}
\end{table}

\subsection{Importance of various ingredients} \label{ingredients}

We can see from these results the importance of some of the specific
ingredients and improvements which we have included in the
calculations. In particular, table (\ref{omctable}) shows the results
obtained for the IA statistical OMC rate using four different methods
of treating the nucleon momentum operator $\vec{q}$.  The most common
way of treating this operator is to replace $\vec{q}$ with
$\frac{\vec{\nu}}{3}$ \cite{Peterson67} (second column) and virtually
all existing OMC calculations used this method.  However, this method
is strictly correct only for S-waves to first order in nucleon
momentum. Using the correct approach discussed in section \ref{sec:ia}
and keeping terms up to ${\mathcal{O}}(\frac{\vec{q}}{\mpp})$
decreases the OMC statistical rate by about $0.5\mathrm\;{s}^{-1}$
(third column) as compared to the results from the
$\frac{\vec{\nu}}{3}$ prescription.  The smallness of the effect is
primarily due to the minute contribution of the P-state wave function
to the trinucleon wave function \footnote{Note that this effect is a
\emph{genuine} effect and not an effect caused by numerical
calculation.  To prove this, a two channel Yamaguchi wave function
consisting solely of S-waves is used to gauge the numerical
uncertainty in wave function integration.  There is an \emph{increase}
of the rate (due only to numerical integration) by
$0.1\;\mathrm{s^{-1}}$, out of a total $1600\;\mathrm{s}^{-1}$, for
the correct approach.  This difference is much smaller than the
difference (stemming from errors in \emph{both} the numerical
integration and the $\frac{\vec{\nu}}{3}$ approach) of the
calculations of the other regular 22-channel wave functions.}.  There
is a difference of about $10\;\mathrm{s}^{-1}$ between using the
$\frac{\vec{\nu}}{3}$ approach (fourth column) and the correct
approach (fifth column) when terms up to
${\mathcal{O}}(\frac{\vec{q}^{2}}{\mpp^{2}})$ are kept.  Although in
percentage terms it is just about $0.6\%$, it is much larger than the
$0.5\;\mathrm{s}^{-1}$ difference between the numbers in the second
and third columns.  This is perhaps to be expected as the
``$\frac{\vec{\nu}}{3}$'' approach cannot be applied to terms higher
than first order in nucleon momentum.

\begin{figure}[htb]
{
\hspace{35mm}
\psfig{figure=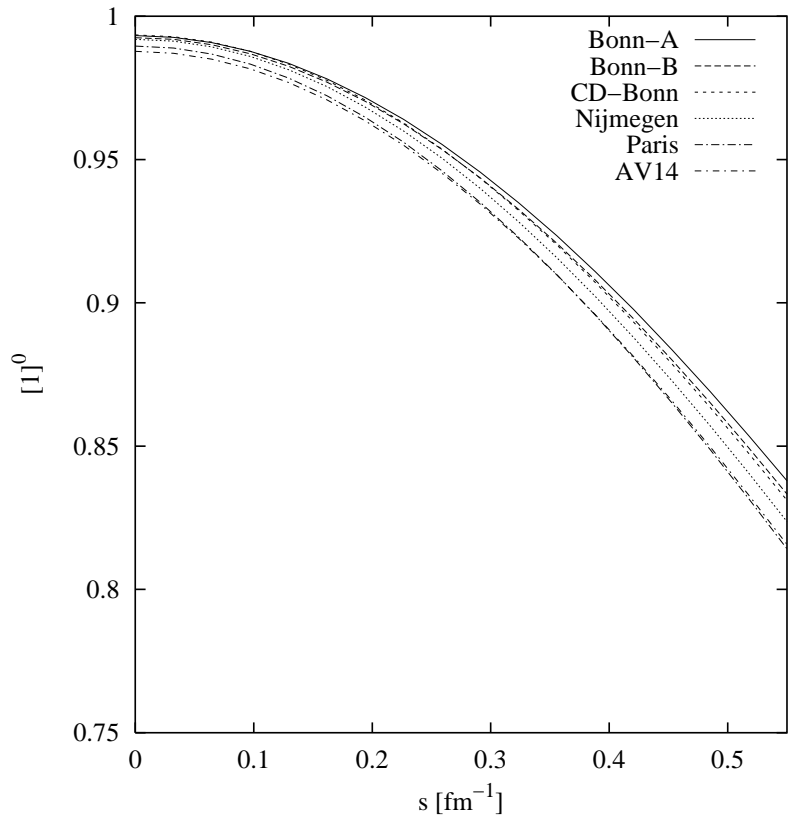,width=110mm,angle=-90}
\vspace{0.75cm}
\caption{Plot of $\ffone$ vs. $s$ for different nuclear 
potentials.\label{iaff_fig1}}
}
\end{figure}

\begin{figure}[htb]
{
\hspace{35mm}
\psfig{figure=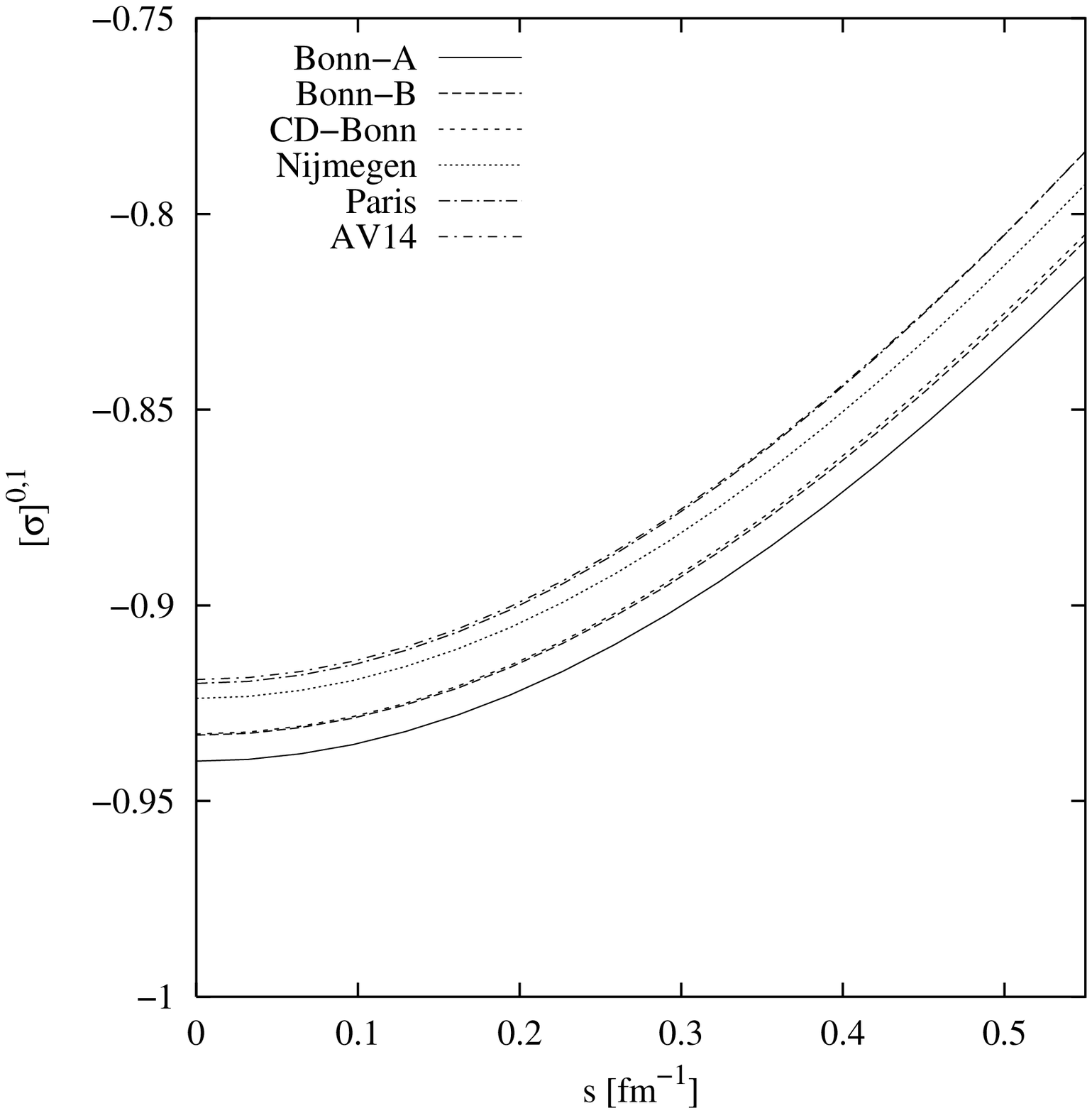,width=110mm,angle=-90}
\vspace{0.75cm}
\caption{Plot of $\fftwo$ vs. $s$.\label{iaff_fig2}}
}
\end{figure}

\begin{figure}[htb]
{
\hspace{35mm}
\psfig{figure=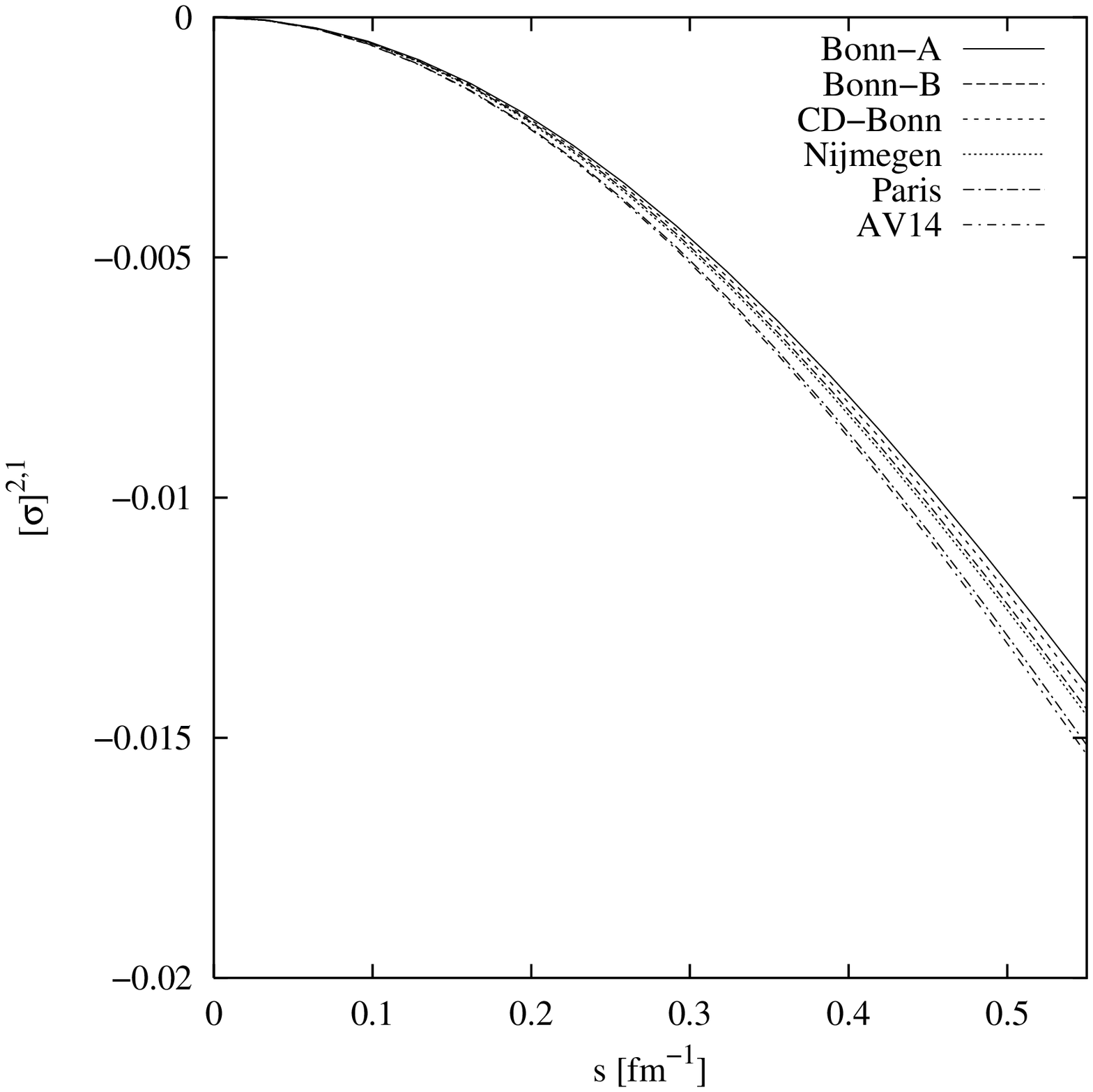,width=110mm,angle=-90}
\vspace{0.75cm}
\caption{Plot of $\ffthree$ vs. $s$.\label{iaff_fig3}}
}
\end{figure}

Table (\ref{omctable}) and Fig. (\ref{rmcfig}) also show the
dependence of both the IA OMC and RMC calculations on various
trinucleon potentials.  In general, the Bonn type potentials seem to
give higher (and perhaps better) RMC and OMC results than the other
potentials.  To analyze this properly, let us take a look at the three
dominant reduced matrix elements, $\ffone$, $\fftwo$ and $\ffthree$,
which are shown as a function of $s$ in Figs. (\ref{iaff_fig1}),
(\ref{iaff_fig2}), and (\ref{iaff_fig3}). All the curves produced by
non-Bonn potentials seem to be a bit separated from the curves of the
Bonn type potentials, especially in the region when $s$ is large.  For
$\ffone$ the problem may be partly associated with the numerical
normalization since $\ffone (s=0)$ should be unity in
principle. However, even though one takes this into account (say, for
example, by scaling the curves so that they agree with
each other at $s=0$) the values of $\ffone$ for the non-Bonn
potentials are still smaller than that of the Bonn potentials, as can
be easily seen from the fact that the fractional deviation of the
reduced matrix element among various wave functions is larger at large
value of $s$.

This seems to have to do with the different binding energies produced
by the Bonn and non-Bonn potentials. The non-Bonn potentials generally
underbind the trinucleon by about 0.5 to 1 MeV. Congleton and Truhlik
\cite{Congleton96} pointed out that\footnote{To see this, expand
$\besselj{0}{\frac{2}{3}sr}$ in polynomials and note that $const > 0$}
$\ffone \sim 1-const \langle r^{2}\rangle s^{2}$ and that $\langle
r^{2}\rangle$ scales like the inverse of binding energy, and thus
argued that one can expect a lower value for $\ffone$ when a
wave function with a lower binding energy is used \footnote{To define
the term ``lower binding energy'', note that appendix (\ref{bind_en})
shows the Bonn potentials have the highest binding energy predictions
among all the wave functions.}. For $\fftwo$, they further argued that
since it is a reduced matrix element for one-body currents (this IA
calculation contains only one-body currents), the Bonn potential's
weak tensor force makes this matrix element large in magnitude.  If
their analysis is correct, this would potentially explain why the Bonn
type potentials consistently give higher results in both OMC and RMC
IA calculations.  The quadratic like curve of $\ffthree$ is obvious if
one notices $\besselj{2}{x}\sim \frac{x^{2}}{15}$ for $x<<1$. Since
$\ffthree$ is the result of S-D overlap, a tri-nucleon wave function
with a larger component of D wave would probably have a larger
magnitude of $\ffthree$ which seems to be the case for the generally
higher D wave component of the non-Bonn potentials (see appendix
(\ref{bind_en})).

Recently Lahiff and Afnan \cite{Lahiff97} suggested that the Bonn-type
potentials might be a better choice than the Nijmegen potential
because the energy dependence of propagators is treated exactly
(during the evaluation of the potential via time-ordered perturbation
theory) in the Bonn type potential but the Nijmegen group removes this
energy dependence.

Thus to summarize, the reason that the Bonn type potentials give a
higher result for both RMC and OMC may primarily be due to its higher
binding energy and possibly its weak tensor force.  The higher D wave
components of those non-Bonn potentials may give a boost to $\ffthree$
but the smallness of $\ffthree$ as compared to the other two would
make its effect on the IA capture rates small.

The effects of the Adler and Dothan terms can also be seen in Fig.
(\ref{rmcfig}).  The upper two curves show the RMC spectrum including
the full $\Delta M$ of of Eq. (\ref{deltaM}) and including only
$\Delta M_1+\Delta M_3$, the part required by GI. There is clearly a
significant difference between these two photon spectra in the EPM.
The photon spectrum with the full Adler and Dothan amplitude is
5-25$\%$ higher for photon momentum starting from 20 MeV than the one
with terms only to ensure GI.

The figure also shows six IA photon spectra for different trinucleon
wave functions using the full set of Adler and Dothan terms plus an IA
photon spectrum with just terms of $\Delta M_1+\Delta M_3$ necessary
to ensure GI. For this latter case a trinucleon wave function from the
Bonn-A potential was used for the calculation.  For the IA case there
is almost no observable change in the photon spectrum produced by
including the full set of Adler-Dothan terms as opposed to just those
required for GI.

If one considers IA and EPM spectra with only the terms from $\Delta
M_1+\Delta M_3$, i.e. only those necessary to ensure GI, the ratio of IA
to EPM RMC total capture rate is similar to that of the corresponding
OMC quantity ($\sim 83-90\%$).  However, as a result of increased
importance of the full set of Adler and Dothan terms in the EPM case,
once these additional terms are included, the IA to EPM RMC capture
rate ratio drops by more than $10\%$ to around $73\%$.

\begin{figure}[htb]  
{  
\hspace{30mm}
\psfig{figure=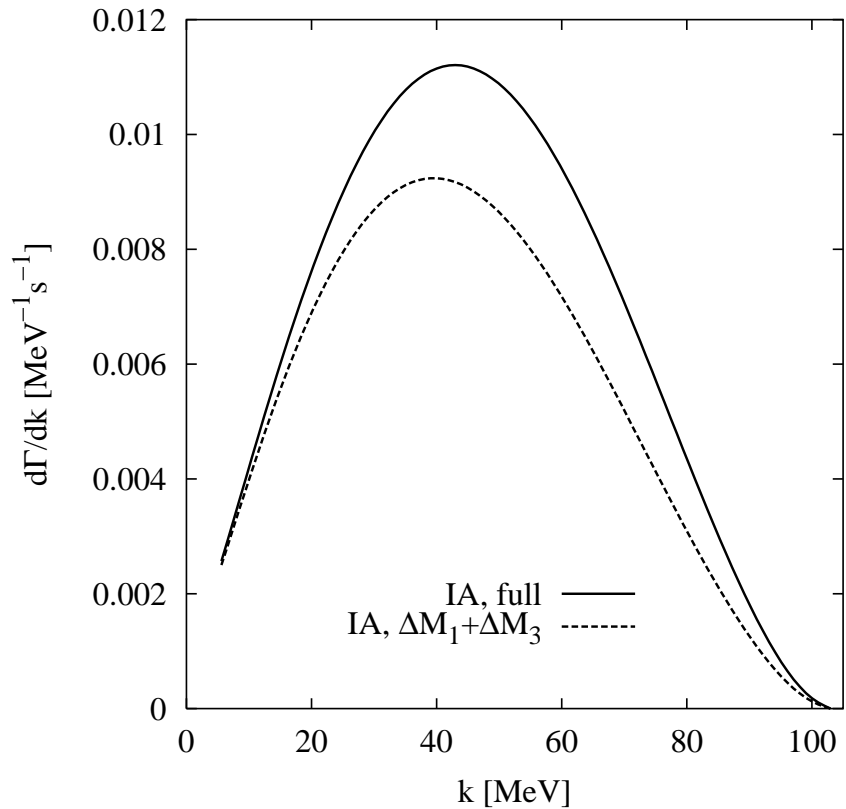,width=110mm,angle=-90}
\vspace{0.75cm}
\enlargethispage*{10pt}
\caption{Effect of increased nucleon radii on the full IA photon spectrum and on the spectrum with only $\Delta M_{1}+\Delta M_{3}$\label{ia.spectrum.radius}}
}
\end{figure}

\begin{figure}[htb]  
{  
\hspace{30mm}
\psfig{figure=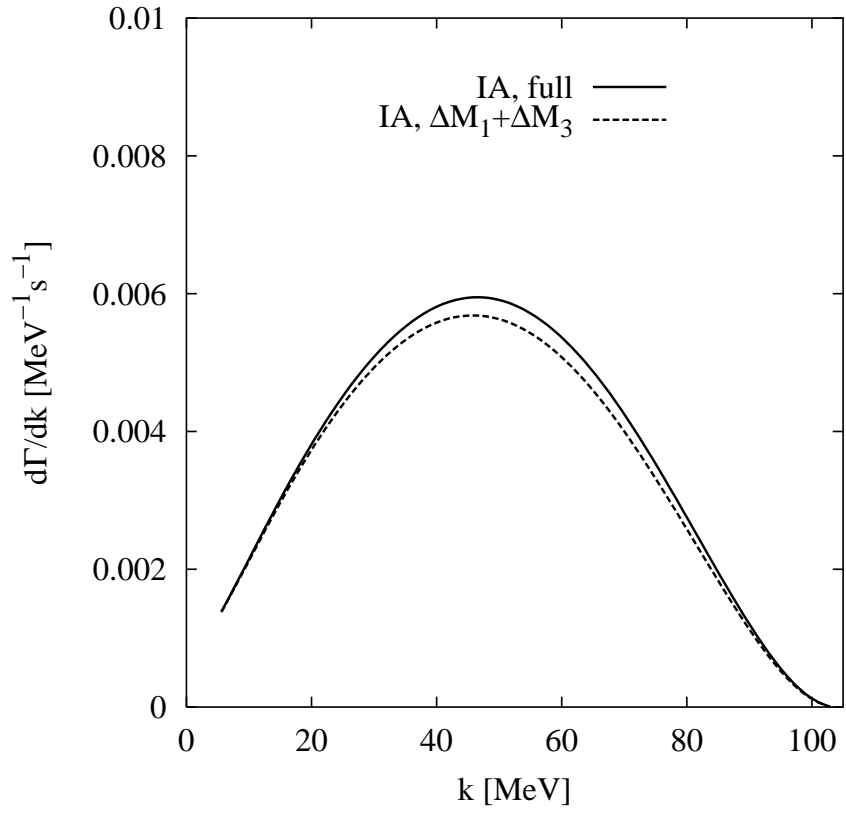,width=110mm,angle=-90}
\vspace{0.75cm}
\caption{Effect of changing the sign of  $g_{A}$ on the two sets of 
IA photon spectrum\label{ia.spectrum.signflip.ga}}
}
\end{figure}

\begin{figure}[htb]  
{  
\hspace{30mm}
\psfig{figure=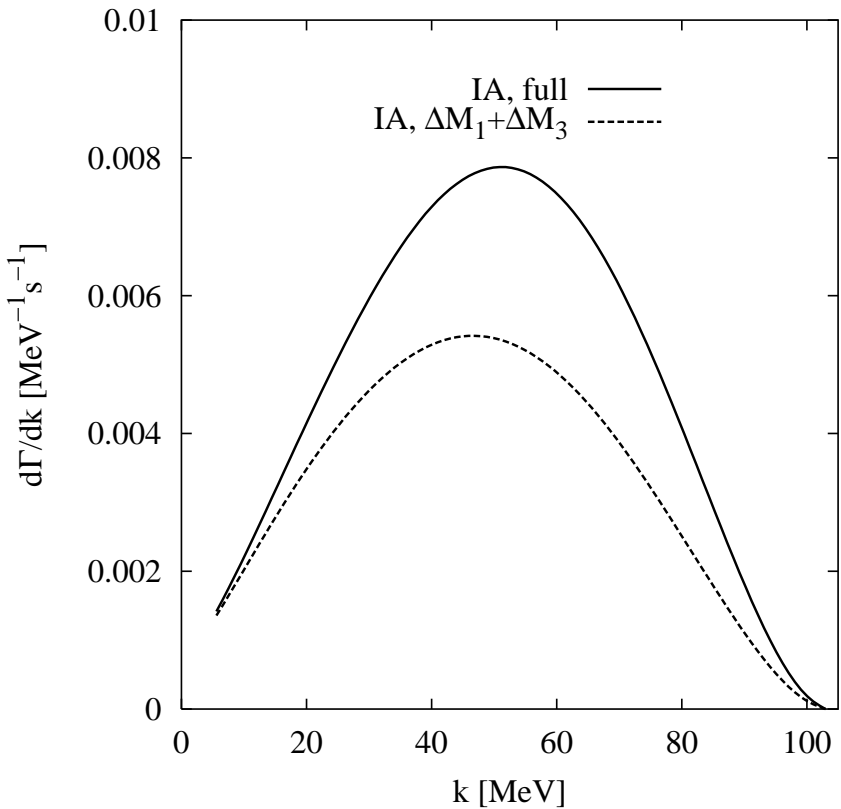,width=110mm,angle=-90}
\vspace{0.75cm}
\caption{Two sets of IA photon spectrum using \emph{both} the increased 
nucleon radii \emph{and} the sign change for $g_{A}$.
\label{ia.spectrum.radius.signflip.ga}}
}
\end{figure}

An obvious reason for this increased sensitivity in the EPM case might
be the more rapidly varying form factors in the nuclear case as
opposed to the nucleon case. However there is also a relative sign
flip between between $F_{A}(0)$ and $g_{A}(0)$ which might play a
role. To investigate this further we have done three things. We have
artificially increased $r_{V}$, $r_{M}$ and $r_{A}$ to
$1.95\;{\mathrm{fm}}$ (a value comparable to various nuclear radii)
and calculated the resulting IA photon spectra using these values
(Fig. (\ref{ia.spectrum.radius})).  We have calculated another set
of photon spectra by flipping the sign of $g_{A}$ in the IA (Fig.
(\ref{ia.spectrum.signflip.ga})).  We also provide the IA photon
spectra when \emph{both} of these effects are present (Fig.
(\ref{ia.spectrum.radius.signflip.ga})).


By looking at Figs. (\ref{ia.spectrum.radius} -
\ref{ia.spectrum.radius.signflip.ga}), one concludes that the IA RMC
calculation would have shown a sensitivity to the additional terms in
the full $\Delta M$ if the nucleon radii were of comparable size to
the nuclear ones, no matter whether the sign of $g_{A}$ is flipped (as
in Fig. (\ref{ia.spectrum.radius.signflip.ga})) or not (as in Fig.
(\ref{ia.spectrum.radius})).  The flipping of sign of $g_{A}$
decreases both spectra (full IA and calculation with only terms to
ensure GI) but the difference between them is relatively small,
provided the various nucleon radii are not artificially increased
(Fig.  (\ref{ia.spectrum.signflip.ga})).

One notes that there are some additional deficiencies of the Adler and
Dothan procedure which may be relevant. In particular,
${\mathcal{O}}(kQ)$ terms in $\Delta M$, which are formally of the
same size as some terms which are kept, cannot be determined uniquely
by GI or CVC and PCAC alone.  As a result, some terms in $\Delta M$
involving derivatives of form factors are missing.  This might be
problematic when those derivatives are large, as in the EPM.

A final remark should be made regarding the comparison of IA and EPM
approaches. A major difference probably originates in the fact that
the IA as described here uses only one body currents. To elucidate
this difference one needs to perform a more detailed investigation of
the interactions of the nucleons using a model, say involving meson
exchanges and adding two (or more)-body currents. These steps would
entail a much more complicated calculation than the one here solely
considering one-body currents.  Congleton and Truhlik
\cite{Congleton96} calculated the MEC contribution to the simpler
problem of OMC by $\mathrm{^{3}He}$ and found that IA+MEC prediction
of the exclusive OMC statistical rate agrees with both the EPM
calculation and experiment \cite{Ackerbauer98}.\footnote{They used a
trinucleon wave function \cite{Kameyama89} that was derived in a
slightly different way than the ones we use.}  
Clearly something similar needs to be done for RMC. Finally one should
note that there are also difficulties with the EPM as applied to RMC
relating to the treatment of intermediate nuclear states, as was
discussed in Ref.\ \cite{Klieb85}.

\subsection{Dependence on $F_P$} \label{Fpdep}

\begin{figure}[htb]
{
\hspace{20mm}
\psfig{figure=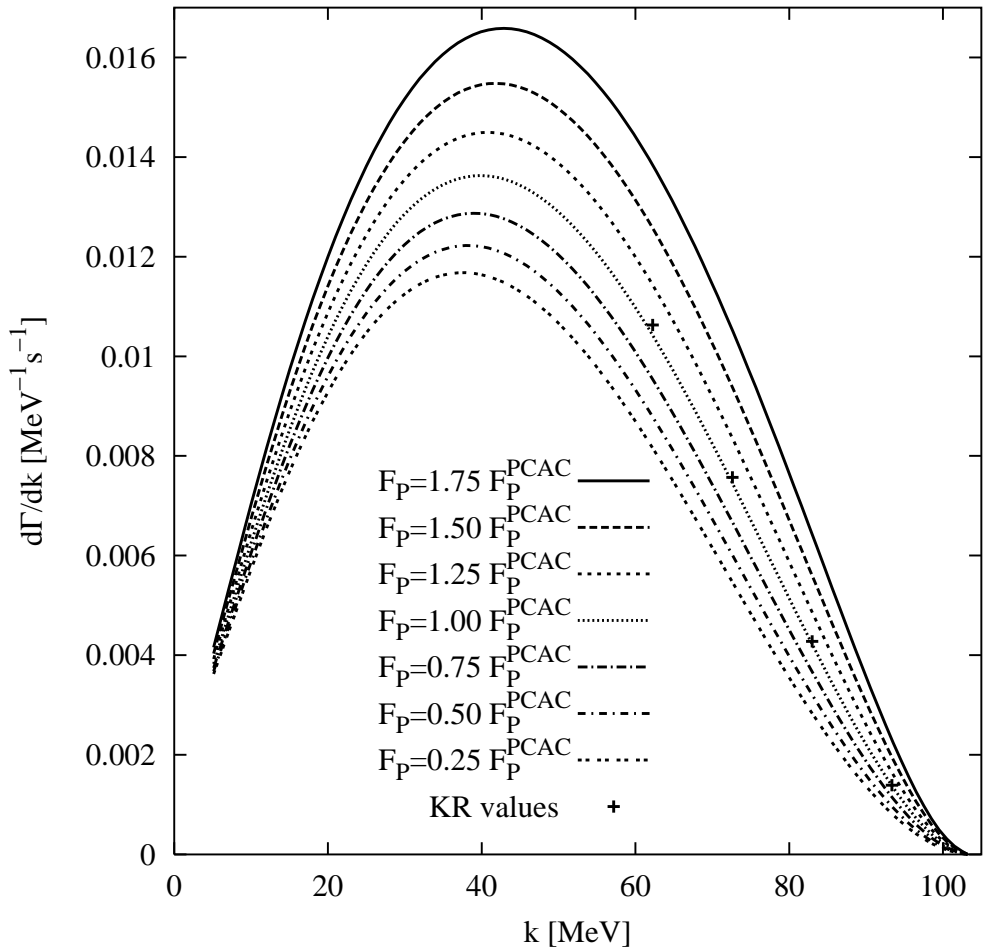,width=130mm,angle=-90}
\vspace{0.75cm}
\caption{The RMC photon spectrum $\frac{d\Gamma_{stat}}{dk}$
calculated in the EPM approach using the full Adler Dothan amplitude
with various values of $F_{P}$, in units of $F_{P}^{PCAC}$ as
determined from Eq. (\protect\ref{pionpole}) with $\varepsilon=0$.
The KR values are those of Klieb and Rood, taken from the relativistic
calculation of Ref. \protect\cite{Kliebthesis} which are not shown in
Ref. \protect\cite{Klieb81} \label{EPM_spectrum}} }
\end{figure}

One of the main motivations for examining RMC is to obtain information
about the induced pseudoscalar coupling constant and hence we have
obtained results for various values of this coupling. Figure
(\ref{EPM_spectrum}) shows the EPM calculation of the RMC photon
spectrum for several different choices of $F_{P}$.  There is an
increase in the total capture rate and a slight shift of the peak of
the spectrum to higher photon energy as one increases $F_{P}$ from
0.25 to 1.75 times its PCAC value.

\begin{figure}[htb]
{
\hspace{20mm}
\psfig{figure=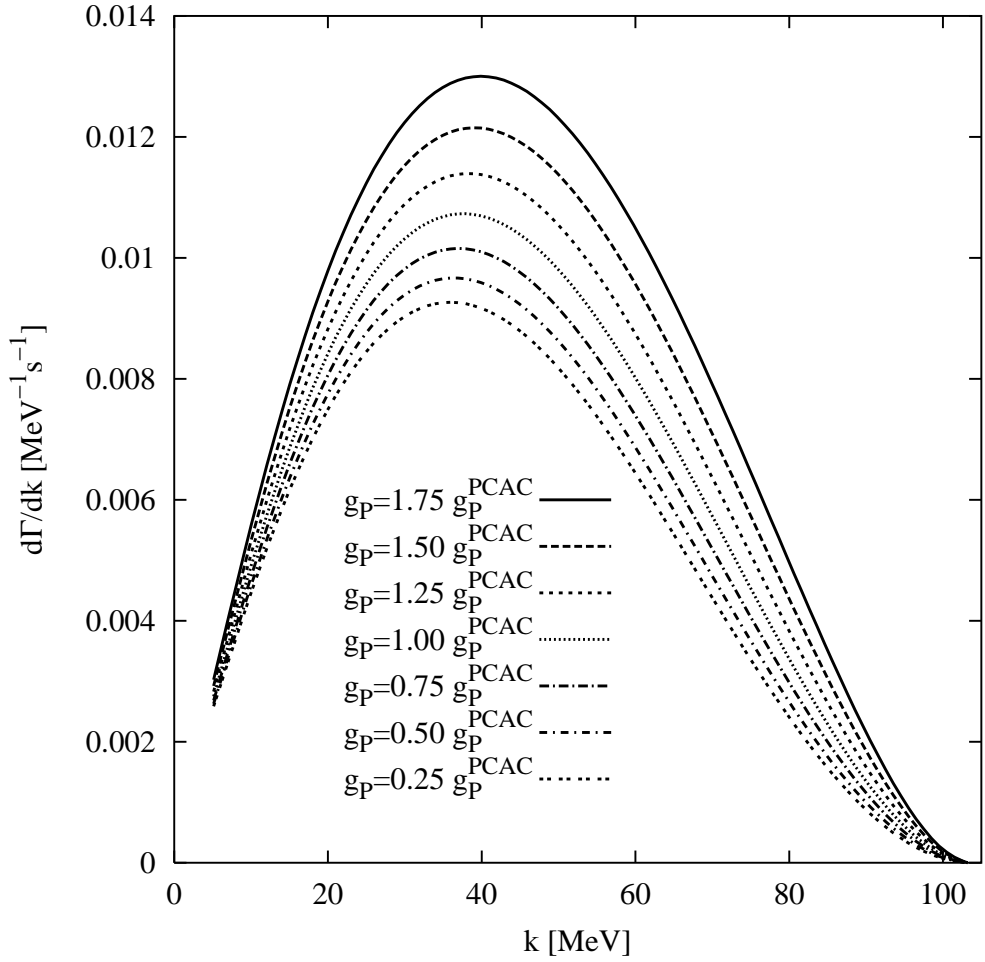,width=130mm,angle=-90}
\vspace{0.75cm}
\caption{Photon spectra from IA calculations for various values of
$g_{P}$, in units of $g_{P}^{PCAC}$ as determined from
Eq. (\protect\ref{pionpole}) with $\varepsilon=0$.  Wave functions
derived from Bonn-A potential are used for the
calculation.\label{bonnafig}} }
\end{figure}

Figure (\ref{bonnafig}) shows the same quantity as Fig.
(\ref{EPM_spectrum}) but for IA calculation using wave functions
derived from the Bonn-A potential. The qualitative features are
essentially the same as for the EPM case, though the absolute
magnitude is different, as was discussed above.

\begin{figure}[htb]
{
\hspace{20mm}
\psfig{figure=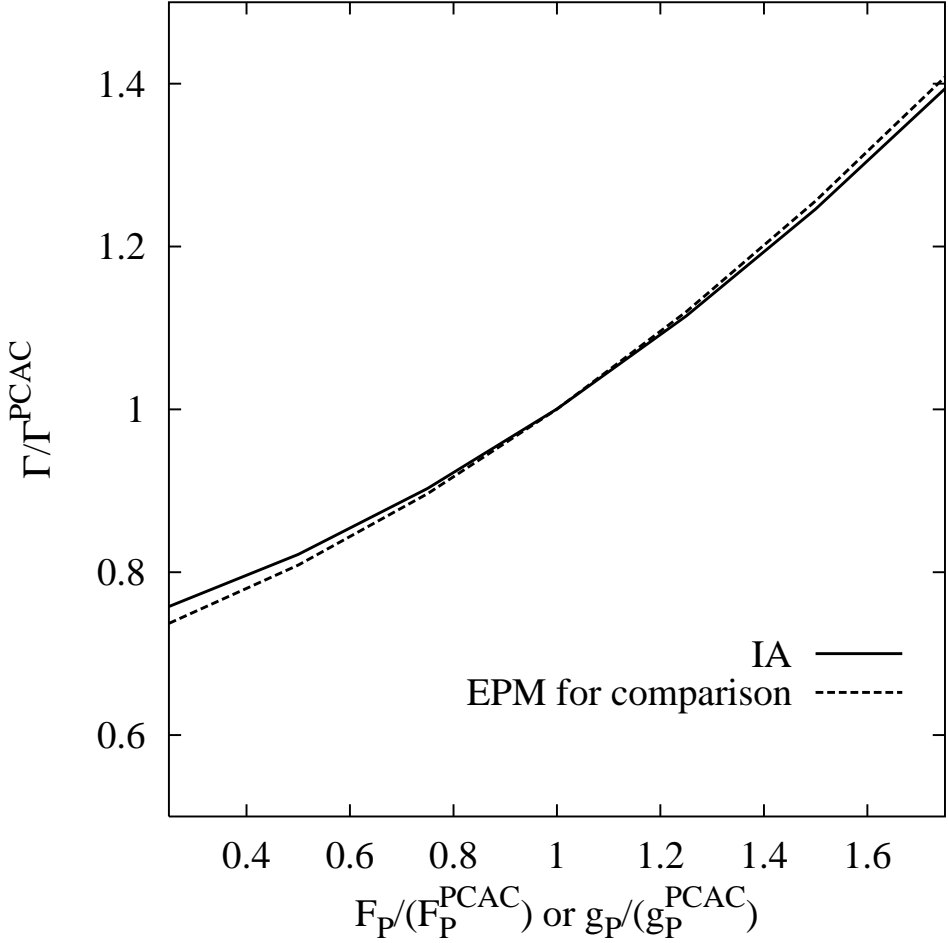,width=130mm,angle=-90}
\vspace{0.75cm}
\caption{Sensitivity of $\Gamma^{rmc}_{stat}(k>57\;\mathrm{MeV})$ to
$F_{P}$ or $g_{P}$.  The Bonn-A potential was used for the IA
calculations and the full $\Delta M$ was included for both IA and EPM
results.\label{total_rmc_vs_gp_ia}} 
}
\end{figure}

Figure (\ref{total_rmc_vs_gp_ia}) shows the sensitivity of the
integrated spectrum, i.e. the RMC capture rate ($k>$57 MeV) with
respect to variation of $F_{P}$ (for the EPM) and $g_{P}$ (for the
IA). The increase in total capture rate as $F_{P}$ or $g_{P}$
increases from 0.25 to 1.75 times its PCAC value is slightly more
rapid in the EPM than in the IA.

\begin{figure}[htb]
{
\hspace{20mm}
\psfig{figure=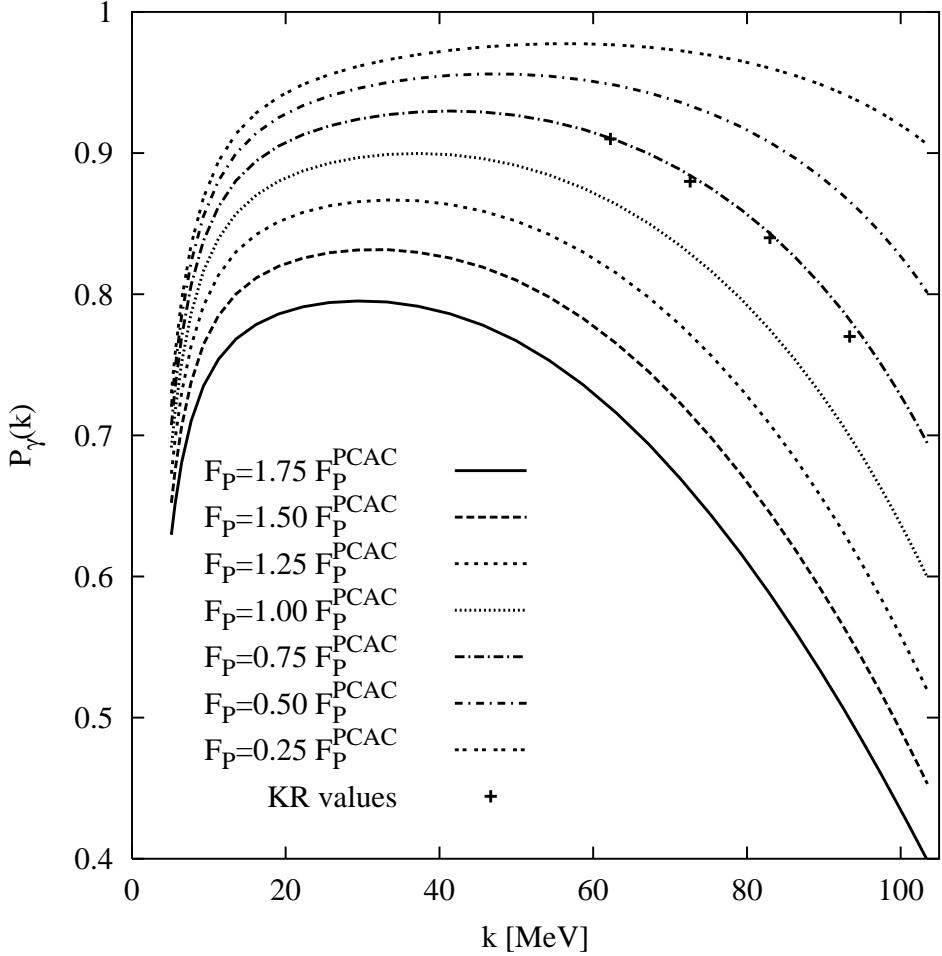,width=130mm,angle=-90}
\vspace{0.75cm}
\caption{\label{EPM_photon_polar}Photon polarization for various
values of $F_{P}$.} 
}
\end{figure}

Another quantity which is sensitive to $F_P$ in RMC and can in
principle be considered, though the experiment is difficult, is the
photon polarization. This is defined as the rate (or spectrum) for
particular photon polarization minus that for the reversed
polarization divided by the sum.  Figure (\ref{EPM_photon_polar})
shows this photon polarization $P_{\gamma}(k)$ using the EPM for
various values of $F_{P}$. Clearly there is a very strong dependence
on $F_{P}$, particularly for the highest energy photons. These photon
polarizations for different values of $F_{P}$ all seem to converge to
a limit as $k\rightarrow 0$.  This is due to the fact that in this
limit the amplitude is determined by soft photon theorems. In the
usual transverse gauge and for initial muon and nucleus at rest, the
leading term in the squared amplitude is
${\mathcal{O}}(\frac{1}{k^{2}})$ and has the form
$|\frac{\dprod{\vec{P_{f}}}{\pol}}{M_{n}k}|^{2}$ (from the diagram
with the final hadron emitting). It is thus independent of the sign of
the photon polarization, which makes $P_{\gamma}(k)\rightarrow 0$ as
$k\rightarrow 0$.

Note that, although we have not calculated it explicitly, similar
sensitivities to $F_{P}$ would be expected in the photon asymmetry
relative to the muon spin.\cite{Fearing75}

\subsection{Comparison with other works}\label{subsec:comp}

The EPM photon spectrum (Fig. (\ref{EPM_spectrum})) and total RMC
rate, given in table (\ref{omctable}), are in good agreement with the
results of Klieb and Rood \cite{Klieb81,Kliebthesis} who obtained a
total rate of $0.814\;\mathrm{ s^{-1}}$ \cite{Kliebthesis} via a
non-relativistically approximated amplitude. Thus the extra terms
included here but not included by Klieb and Rood seem not to
contribute significantly to the photon spectrum.  The agreement of the
EPM OMC rate of $1503\;\mathrm{s}^{-1}$ with Congleton and Fearing's
\cite{Congleton93} is also good, though this agreement is perhaps not
surprising given that the values of the form factors used here, and
the basic approach, are almost the same as theirs.

The IA calculation of OMC rate using wave functions derived from Paris
or AV14 potentials agrees with that of Klieb and Rood. However, the
two IA RMC spectra are about $4\%$ lower\footnote{This already takes
into account the fact that they took the value of $C=0.965$ while
$C=0.9788$ is used in this work.} than theirs.  At first it looks a
bit contradictory that more or less the same IA OMC rates, but
slightly lower RMC results, are obtained in this work, but a closer
look reveals that Klieb and Rood used a lot of approximations
evaluating the reduced matrix elements for the RMC spectrum which they
did not use for OMC.

In particular instead of evaluating the reduced matrix elements
directly in terms of $s$, they expressed $\hat{s}$ in terms of an
infinite sum of spherical harmonics of $\hat{\nu}$ and $\hat{k}$ and
imposed an artificial cutoff on this expansion.  They also did not
fully square the resulting matrix element.  Only products of any two
of the most dominant terms and products of one dominant and one small
term were considered.  In expanding the plane wave
$\exp(i\dprod{\vec{s}}{\vec{r}})$, they only included the term having
$\besselj{0}{\nu r}\besselj{0}{kr}$ and they used this approximation
as the premise in deriving several relationships between various
reduced matrix elements for RMC.  They also do not seem to have
expanded the spinor normalization factors
$\sqrt{\frac{\mpp+\kpao}{2\kpao}}$ (for the neutron spinor) and
$\sqrt{\frac{\mpp+\kao}{2\kao}}$ (for the proton spinor) quite
consistently.  None of these approximations were used here.  Also some
terms in $\Delta M$ that are present here but not included by Klieb
and Rood tend to decrease the resulting photon spectrum a bit.  Given
these differences, the agreement within 4\% for the rates is quite
satisfactory.

The EPM photon polarization $P_{\gamma}(k)$ obtained in this work does
differ from that of Klieb and Rood quite significantly (see Fig.
\ref{EPM_photon_polar}).  The reason behind this is that while the
non-relativistically reduced amplitude used by Klieb and Rood produces
the correct spectrum within a few percent, it cannot produce
$P_{\gamma}(k)$ accurately.  Fearing \cite{Fearing75} noted that the
first order contribution to $P_{\gamma}(k)$ actually comes from
${\mathcal{O}}(\frac{1}{\mtri^{2}})$ terms in the squared Hamiltonian.
Klieb and Rood apparently compromised $P_{\gamma}(k)$'s accuracy by
truncating many ${\mathcal{O}}(\frac{1}{\mtri^{2}})$ terms when they
squared their already non-relativistically reduced amplitude.

\section{Summary and conclusion}\label{sec:summary} 

We have performed a theoretical calculation of the process
$\mathrm{^{3}He}+\mu^{-}\rightarrow\mathrm{^{3}H}+\gamma+\nu_{\mu}$
using two separate approaches, the elementary particle model and the
impulse approximation.  Our calculation contains a number of
improvements over the previous ones, namely: 1) The full Adler and
Dothan amplitude is used for both the EPM and IA calculations. 2)
Better momentum space wave functions from various nuclear potentials
are employed for the IA calculation. 3) The non-relativistic reduction
of the IA amplitude contains second order nucleon momentum terms for
coefficients of $g_{P}$. 4) The nucleon momentum terms in the IA are
treated exactly without using the common $\frac{\vec{s}}{3}$ approach.

In general our results agree well, when comparisons can be made, with
the older calculations of Klieb and Rood \cite{Klieb81,Kliebthesis}.
In particular using the EPM model approach, the RMC statistical rate
obtained in this work agrees with that of Klieb and Rood.The photon
polarization $P_{\gamma}(k)$ disagrees significantly with their
results, which probably is a consequence of the fact that they
truncated a lot of ${\mathcal{O}}(\frac{1}{\mtri^{2}})$ terms in the
squared amplitude which contribute significantly to $P_{\gamma}(k)$.
The IA OMC results derived from wave functions of the non-Bonn
potentials roughly agree with Klieb and Rood's but the RMC photon
spectra from the same wave functions are slightly lower than Klieb and
Rood's.  This seems to have to do with the fact that they made some IA
RMC specific approximations in evaluating their photon spectrum.

We summarize our results as follows. As expected there is a strong
dependence of the results on the value of the induced pseudoscalar
coupling constant $F_P$ or $g_P$. There is a slight dependence of the
IA calculations on nuclear potentials.  The dependence can possibly be
accounted for by the difference in the three body binding energy
resulting from the different potentials, by details of the nuclear
potentials such as stronger/weaker tensor force, etc., and by
differing partial wave characteristics of the resulting trinucleon
wave functions.

There is quite a significant difference between the EPM and IA RMC
calculations.  A first look at the spectra would suggest that the
difference is caused by the fact that the EPM calculation is much more
sensitive than the IA to the ${\mathcal{O}}(k,k^{2})$ terms in $\Delta
M$ which are larger in the EPM because of the much larger magnitude of
the various nuclear radii than their nucleon counterparts. This hints
at a poor convergence of these Adler and Dothan terms in the EPM and
suggests that the higher order pieces which cannot be calculated via
the Adler-Dothan procedure might be important. Further problems with
the EPM have been discussed before in Ref. \cite{Klieb85}.  Probably
the most important effect contributing to this difference however are
the meson exchange corrections. These accounted for the difference
between IA and EPM in the OMC case \cite{Congleton96}. Such a
calculation will be much more complicated for RMC, but is clearly
needed to fully understand the differences between the IA and EPM
approaches.

\section*{Acknowledgments}

This work was supported in part by the Natural Sciences and
Engineering Research Council of Canada.  

\appendix

\section{$M(\Phfour,\Phefour,\mtri)$ in the EPM}
Using the Dirac representation of the $\gamma$ matrices, $M(\Phfour,
\Phefour, \mtri)$ (up to a constant phase factor) can be written in 
the form below which operates on the product space of the leptonic 
and hadronic spinors.
\begin{eqnarray}
M(\Phfour, \Phefour,
\mtri)&=&{\mathrm{N^{\prime}}}(1-\dprod{\sml}{\hat{\nu}})
\Big{\{}f_{1}\dprod{\sml}{\pol}+ f_{2}\dprod{\sm}{\pol}+
if_{3}\dprod{\cprod{\pol}{\sml}}{\sm}+
\frac{f_{4}}{2m}\dprod{\sml}{\pol}\dprod{\sm}{\vec{s}}+ \nonumber\\
&&\frac{f_{6}}{2m}\dprod{\vec{\nu}}{\pol}+
i\frac{f_{7}}{2m}\dprod{\cprod{\vec{s}}{\pol}}{\sm}+
\frac{f_{8}}{2m}\dprod{\sml}{\vec{s}}\dprod{\sm}{\pol}+
\frac{f_{9}}{2m}\dprod{\sml}{\sm}\dprod{\vec{\nu}}{\pol}+ \nonumber\\
&&\frac{f_{10}}{4m^{2}}\dprod{\sm}{\vec{s}}\dprod{\vec{\nu}}{\pol}+
i\frac{f_{11}}{4m^{2}}\dprod{\cprod{\vec{s}}{\vec{\nu}}}{\sm}
\dprod{\sml}{\pol}+
\frac{f_{12}}{4m^{2}}\dprod{\sm}{\vec{\nu}}\dprod{\vec{\nu}}{\pol}+
\frac{f_{13}}{2m}\dprod{\sm}{\vec{\nu}}\dprod{\sml}{\pol}+ \nonumber\\
&&i\frac{f_{14}}{4m^{2}}\dprod{\cprod{\vec{s}}{\pol}}{\sm}\dprod{\sml}
{\vec{s}}+
i\frac{f_{15}}{8m^{3}}\dprod{\cprod{\vec{s}}{\vec{\nu}}}{\sm}
\dprod{\vec{\nu}}{\pol}\Big{\}}
\end{eqnarray}
where $\sml$ is the leptonic spin matrix and $\sm$ is the hadronic
spin matrix.  $\mathrm{N}^{\prime}$ is the spinor normalization factor
which equals $\frac{1}{2}\sqrt{\frac{\Pho+\mtri}{2\Pheo}}$ (the factor
$\frac{1}{2}$ comes from the normalizations of the photon and the
neutrino).  The infrared divergent part comes from $\frac{1}{\zeta}$
where $\frac{1}{\zeta}=\frac{1}{2(\Pho
k+k^{2}+\dprod{\vec{\nu}}{\vec{k}})}$, a term of
${\mathcal{O}}(\frac{1}{k})$.  
\begin{eqnarray}
f_{1}&=&\frac{F_{V}^{H}}{\Pho+\mtri}\Big{\{}-(1+\frac{k}{\mtri}+
\frac{\Pho}{\mtri}+\frac{\mtri
k+k^{2}+2\dprod{\vec{\nu}}{\vec{k}}+\Pho
k}{\zeta})+\lambda(\frac{\nu}{\mtri}-\frac{\nu k}{\zeta})-
{\kappa_{i}}\frac{k-\lambda\nu}{2\mtri}+ \nonumber\\
&&{\kappa_{f}}(\frac{k}{2\mtri}\frac{2k\Pho-2\dprod{\vec{\nu}}
{\vec{k}}}{\zeta}-\lambda(\frac{\nu
k^{2}+\Pho\nu k}{\mtri\zeta}))\Big{\}}+ \nonumber\\
&&{F_{V}^{L}}\Big{\{}\frac{1}{2}(\lambda-1)\frac{1}{m}(1+
\frac{\nu+k}{\Pho+\mtri})\Big{\}}+
\nonumber\\
&&\frac{F_{V}^{\prime}}{\Pho+\mtri}\Big{\{}2(\nu^{2}-s^{2}+
\dprod{\vec{\nu}}{\vec{k}})-2(\Pho+\mtri)k+(\kappa_{i}-
\kappa_{f})\frac{\lambda\nu
k^{2}+k\dprod{\vec{\nu}}{\vec{k}}}{\mtri}\Big{\}}+ \nonumber\\
&&\frac{F_{M}^{H}}{\Pho+\mtri}\Big{\{}(\lambda+1)\frac{\nu}{2\mtri}
(1+\frac{\nu}{\mtri})
+(\frac{\Pho-k}{2\mtri})(\frac{\nu-m}{\mtri}) -\frac{k(\nu-m)}{2\zeta}
+\frac{\lambda\nu}{2\mtri}(\frac{\Pho-k-m}{\mtri})- \nonumber\\
&&\frac{m}{2\mtri}(1+\frac{-\Pho
k-2\dprod{\vec{\nu}}{\vec{k}}-k^{2}}{\zeta})
-\frac{\nu}{2\mtri}\frac{2\dprod{\vec{\nu}}{\vec{k}}-(1-\lambda)\nu
k+(\lambda+1)(\Pho k+k^{2})}{\zeta}- \frac{\lambda\nu
k}{2\zeta}(1-\frac{m}{\mtri})+ \nonumber\\
&&\frac{\lambda\kappa_{i}\nu}{4\mtri}(1+\frac{\Pho+\nu-m}{\mtri})
+\frac{\kappa_{i}(m-\nu)k}{4\mtri^{2}}+
\kappa_{f}\Big{(}-\frac{\lambda\nu
k^{2}}{2\mtri\zeta}(1+\frac{\Pho+\nu-m}{\mtri})+ \nonumber\\
&&\frac{mk}{4\zeta}(1-\frac{\Phos}{\mtri^{2}}+\frac{2\dprod{\vec{\nu}}
{\vec{k}}-s^{2}-2\Pho
k}{\mtri^{2}})- \frac{\nu k}{4\zeta}(1-\frac{2\nu}{\mtri}-\frac{2(\Pho
k+\nu
k-\dprod{\vec{\nu}}{\vec{k}}+{\nu}\Pho)+s^{2}+\Phos}{\mtri^{2}})-
\nonumber\\ &&\frac{\lambda\nu k}{4\zeta}(1+\frac{2\Pho}{\mtri}
+\frac{2(\Pho\nu-\Pho
m-\dprod{\vec{\nu}}{\vec{k}})+\Phos+s^{2}}{\mtri^{2}})\Big{)}\Big{\}}+
\nonumber\\
&&\frac{F_{M}^{L}}{\Pho+\mtri}\Big{\{}(\lambda-1)(-\frac{\nu+k}{4\mtri}+
\frac{(\nu+k)^{2}-s^{2}}{4m\mtri})\Big{\}}+
\nonumber\\
&&\frac{F_{M}^{\prime}}{\Pho+\mtri}\Big{\{}\frac{m-\nu}{\mtri}
(\dprod{\vec{\nu}}{\vec{k}}+\lambda\nu
k)\Big{\}}+ \nonumber\\
&&\frac{F_{A}^{H}}{\Pho+\mtri}\Big{\{}-\lambda(1+\frac{\Pho+k}{\mtri}
(1-\frac{\mtri
k}{\zeta})-\frac{\mtri
k}{\zeta})+\frac{\nu}{\mtri}-\frac{k\nu}{\zeta}-\frac{\lambda
\kappa_{i}}{2}(1+\frac{\Pho}{\mtri})+ \nonumber\\
&&{\lambda\kappa_{f}}(\frac{2k(\Pho+k)+\mtri
k}{2\zeta}+\frac{\Pho}{2\mtri}\frac{k(2k+\Pho)}{\zeta}+\frac{k}{2\mtri}
\frac{s^{2}-2\dprod{\vec{\nu}}{\vec{k}}}{\zeta})
-{\kappa_{f}}(\frac{k}{2\mtri}\frac{2\nu(\Pho+k)}{\zeta}+
\frac{k\nu}{\zeta})
\Big{\}} \nonumber\\
f_{2}&=&\frac{F_{V}^{H}}{\Pho+\mtri}\Big{\{}-\frac{\lambda(2+
\kappa_{i})}{2}(1+\frac{\Pho}{\mtri})+\frac{\lambda
k(\mtri+\Pho)}{\zeta}+{\lambda\kappa_{f}}((1+\frac{\Pho}{2\mtri})
\frac{\Pho
k}{\zeta} +\frac{\mtri k}{2\zeta}+\frac{ks^{2}}{2\mtri\zeta})\Big{\}}+
\nonumber\\
&&F_{V}^{\prime}\Big{\{}\lambda(\kappa_{i}-\kappa_{f})k(\frac{\nu+k}
{\mtri})\Big{\}}+
\nonumber\\
&&\frac{F_{M}^{H}}{\Pho+\mtri}\Big{\{}-\frac{k}{2\mtri}(1+\frac{\Pho}
{\mtri}+\frac{\mtri
k}{\zeta})+\frac{2\lambda+\kappa_{i}}{4}(-\frac{k}{\mtri}(1+\frac{\Pho}
{\mtri}))+\frac{\lambda(2+\kappa_{i})}{4}\frac{s^{2}}{\mtri^{2}}+
\lambda(\frac{\mtri
k-k^{2}}{2\zeta}+ \nonumber\\
&&\frac{k}{\mtri}\frac{2s^{2}-\Phos-(1+\lambda)\Pho k}{2\zeta})+
\kappa_{f}\Big{(}\frac{\lambda
k}{4\mtri}(\frac{s^{2}}{\zeta}(1-\frac{\Pho-k}{\mtri})-
\frac{\dprod{\vec{\nu}}{\vec{k}}+\nu^{2}}{\zeta}(1+\frac{\Pho}{\mtri}))
-\frac{k^{2}}{4\zeta}(1+\frac{\Pho}{\mtri}(2+\frac{\Pho}{\mtri})+
\nonumber\\ &&\lambda(1-\frac{\Phos}{\mtri^{2}}))\Big{)}\Big{\}}+
\frac{F_{A}^{H}}{\Pho+\mtri}\Big{\{}-1-\frac{\Pho}{\mtri}-\frac{(\Pho
+\mtri)k}{\zeta}+
{\kappa_{f}}(\frac{k}{2\mtri}\frac{\Phos+s^{2}}{\zeta}-\frac{\mtri
k}{\zeta})\Big{\}}+ \nonumber\\
&&F_{A}^{L}\Big{\{}\frac{\lambda-1}{2m}\Big{\}}+F_{A}^{\prime}
\Big{\{}-2(\nu+k)\Big{\}}+
\nonumber\\
&&\frac{F_{P}^{H}}{\Pho+\mtri}\Big{\{}(-1-\frac{\Pho}{\mtri}+
\frac{(\mtri+\Pho)k}{\zeta})+{\kappa_{f}}(\frac{k}{2\mtri}(-
\frac{\Phos+s^{2}}{\zeta})+
\frac{\mtri k}{\zeta})\Big{\}} \nonumber\\
f_{3}&=&\frac{F_{V}^{H}}{\Pho+\mtri}\Big{\{}1+\frac{\Pho}{\mtri}-
\frac{(\mtri+\Pho)k}{\zeta}+\kappa_{f}(-\frac{\mtri
k}{2\zeta}+ \frac{k}{2\mtri}(\frac{\Phos+s^{2}}{\zeta}))\Big{\}}+
\nonumber\\
&&\frac{F_{M}^{H}}{\Pho+\mtri}\Big{\{}\frac{2m-(2+\lambda(2+
\kappa_{i}))\nu}{4\mtri}(1+\frac{\Pho}{\mtri})+
\frac{k(m-(1+\lambda)\nu)}{2\zeta}(1+\frac{\Pho}{\mtri})+ \nonumber\\
&&{\kappa_{f}}\frac{k(\nu-m)}{4\zeta}(\frac{s^{2}+\Phos}{\mtri^{2}}-1)-
{\lambda\kappa_{f}}\frac{k\nu}{4\zeta}(1+\frac{\Pho}{\mtri})^{2}\Big{\}}+
F_{M}^{L}\Big{\{}\frac{1}{2\mtri}\Big{\}}+
\frac{F_{A}^{H}}{\Pho+\mtri}\Big{\{}\frac{\lambda(2+\kappa_{i})}{2}(1+
\frac{\Pho}{\mtri})+
\nonumber\\ &&{\lambda}(\frac{k(\mtri+\Pho)}{\zeta})+
{\lambda\kappa_{f}}(\frac{k(\mtri+2\Pho)}{2\zeta}+
\frac{k}{\mtri}\frac{\Phos+s^{2}}{2\zeta})\Big{\}}+F_{A}^{L}\Big{\{}
\frac{\lambda-1}{2m}\Big{\}}
\nonumber\\
f_{4}&=&\frac{F_{V}^{H}}{\Pho+\mtri}\Big{\{}\lambda(2+\kappa_{i})
\frac{m}{\mtri}-\frac{2\lambda
mk}{\zeta}(1-\kappa_{f}\frac{\Pho}{\mtri})\Big{\}}+\frac{F_{M}^{H}}
{\Pho+\mtri}\Big{\{}{\lambda(2+\kappa_{i})}\frac{m(\nu-m)}
{2\mtri^{2}}+{\lambda}\frac{m(m-\nu)}{\zeta}\frac{k}{\mtri}+
\frac{m\nu}{\mtri^{2}}+
\nonumber\\ &&{\lambda\kappa_{f}}\frac{m(\nu-m)}{\mtri^{2}}\frac{\Pho
k}{\zeta}\Big{\}}+
\frac{F_{A}^{H}}{\Pho+\mtri}\Big{\{}\frac{2m}{\mtri}(1+\frac{\mtri
k}{\zeta})+{\kappa_{f}}\frac{2m}{\mtri}\frac{(\Pho+\mtri)k}{\zeta}
\Big{\}}+F_{A}^{L}\{\frac{1-\lambda}{\Pho+\mtri}\Big{\}}+
\nonumber\\ &&F_{A}^{\prime}\Big{\{}4m+\frac{4mk}{\Pho+\mtri}\Big{\}}+
F_{P}^{L}\Big{\{}\frac{\lambda+1}{\Pho+\mtri}\Big{\}} \nonumber\\
f_{6}&=&\frac{F_{V}^{H}}{\Pho+\mtri}\Big{\{}\frac{2m}{\mtri}(1+
\frac{\lambda}{2}(2+\kappa_{i}))+(1-\lambda)\frac{2mk}{\zeta}+
\frac{4m(\Pho+\mtri+\nu)}{\zeta}+\kappa_{f}\frac{2m}{\mtri}
\frac{k\nu-\lambda(\Pho
k+k^{2})}{\zeta}\Big{\}}+ \nonumber\\
&&\frac{F_{V}^{\prime}}{\Pho+\mtri}\Big{\{}4m(\nu+k+\mtri+\Pho)-
(1+\lambda)(\kappa_{f}-\kappa_{i})\frac{2mk^{2}}{\mtri}\Big{\}}+
\nonumber\\
&&\frac{F_{M}^{H}}{\Pho+\mtri}\Big{\{}(1+\lambda)\frac{m}{\mtri}(1+
\frac{\nu-k+\Pho}{\mtri}-\frac{k^{2}}{\zeta})+\frac{m(\mtri-k)}
{\zeta}+\frac{m}{\mtri}(\frac{\lambda
k(m-\nu-\mtri-\Pho)}{\zeta}+ \nonumber\\
&&\frac{k\nu-2m\nu-k\Pho+s^{2}-2\dprod{\vec{\nu}}{\vec{k}}-\Phos}
{\zeta})-\lambda\frac{m^{2}}{\mtri^{2}}+\lambda\kappa_{i}\frac{m}
{2\mtri}(1+\frac{\Pho+\nu-m}{\mtri})-\kappa_{i}\frac{m}{2\mtri}
\frac{k}{\mtri}-
\nonumber\\
&&\kappa_{f}\frac{mk}{2\zeta}(1-\frac{2\nu}{\mtri}+
\frac{2m\nu-2k\nu-2\Pho
k-\Phos+2\dprod{\vec{\nu}}{\vec{k}}-s^{2}-2\Pho\nu}{\mtri^{2}})+
\lambda\kappa_{f}(-\frac{mk}{2\zeta}((1+\frac{\Pho}{\mtri})^{2}-
\nonumber\\ &&\frac{2\dprod{\vec{\nu}}{\vec{k}}-s^{2}-2\Pho k-2\nu
k+2mk+2m\Pho-2\nu\Pho}{\mtri^{2}}+\frac{2k}{\mtri}))\Big{\}}+
F_{M}^{L}\Big{\{}\frac{1}{\Pho+\mtri}\frac{m}{\mtri}\Big{\}}+
\nonumber\\
&&\frac{F_{M}^{\prime}}{\Pho+\mtri}\Big{\{}-\frac{2m}{\mtri}
(\dprod{\vec{\nu}}{\vec{k}}+m\nu)+\lambda\frac{2mk}{\mtri}
(m-\nu)\Big{\}}+
\nonumber\\
&&\frac{F_{A}^{H}}{\Pho+\mtri}\Big{\{}(1+\lambda)(\frac{2m}
{\mtri}-(1+\kappa_{f})\frac{2mk}{\zeta})+\lambda\kappa_{i}
\frac{m}{\mtri}-\kappa_{f}(\frac{2m}{\mtri}\frac{k^{2}+k\Pho}{\zeta}-
\frac{2\lambda
m}{\mtri}\frac{\nu k}{\zeta})\Big{\}}+ \nonumber\\
&&\frac{F_{P}^{H}}{\Pho+\mtri}\Big{\{}-\lambda\frac{m}{\mtri}
(2+\kappa_{i})+(\kappa_{f}+1)\lambda\frac{2mk}{\zeta}\Big{\}}
\nonumber\\
f_{7}&=&\frac{F_{V}^{H}}{\Pho+\mtri}\Big{\{}\frac{2m}{\mtri}+
\frac{2mk}{\zeta}(1+\kappa_{f}(1+\frac{\Pho}{\mtri}))\Big{\}}+
F_{V}^{L}\Big{\{}\frac{1-\lambda}{\Pho+\mtri}\Big{\}}+
\frac{F_{V}^{\prime}}{\Pho+\mtri}\Big{\{}{4m(\nu+k)}+
{2(\kappa_{i}-\kappa_{f})}\frac{k}{\mtri}
\nonumber\\
&&m(\nu+k)\Big{\}}+\frac{F_{M}^{H}}{\Pho+\mtri}\Big{\{}(1+
\lambda)(\frac{m}{\mtri}+\frac{mk}{\zeta})(1+\frac{\Pho}{\mtri})-
\frac{2+\kappa_{i}}{2}\frac{m}{\mtri}\frac{k}{\mtri}-
\frac{m\mtri}{\zeta}(1-\frac{\Phos-s^{2}-k^{2}}{\mtri^{2}})+
\nonumber\\
&&\frac{\kappa_{f}}{2}\frac{mk}{\zeta}(1-\frac{k}{\mtri}-
\frac{\Pho(\Pho-k)+s^{2}}{\mtri^{2}})+
\frac{\kappa_{f}\lambda}{2}\frac{mk}{\zeta}(1+\frac{\Pho}{\mtri})^{2}+
{\kappa_{i}\lambda}\frac{m}{2\mtri}(1+\frac{\Pho}{\mtri})\Big{\}}+
\nonumber\\
&&\frac{F_{M}^{L}}{\Pho+\mtri}\Big{\{}\frac{1-\lambda}{2}(1+
\frac{\Pho+\nu+k}{\mtri})+(\lambda+1)\frac{m}{2\mtri}\Big{\}}+
\frac{F_{A}^{H}}{\Pho+\mtri}\Big{\{}{\lambda(2+\kappa_{i})}
\frac{m}{\mtri}+{\lambda}(-\frac{2mk}{\zeta}(1-
\frac{\kappa_{f}\Pho}{\mtri}))\Big{\}}+
\nonumber\\
&&\frac{F_{P}^{H}}{\Pho+\mtri}\Big{\{}-{\lambda}\frac{m}{\mtri}(2+
\kappa_{i})+
{\lambda}\frac{2mk}{\zeta}(1-\frac{\kappa_{f}\Pho}{\mtri})\Big{\}}
\nonumber\\
f_{8}&=&\frac{F_{V}^{H}}{\Pho+\mtri}\Big{\{}{\lambda}\frac{m}{\mtri}
(2+\kappa_{i})+{\lambda}\frac{2mk}{\zeta}(1-\frac{\kappa_{f}\Pho}
{\mtri})\Big{\}}+F_{V}^{L}\Big{\{}\frac{\lambda-1}{\Pho+\mtri}\Big{\}}+
\nonumber\\
&&F_{V}^{\prime}\Big{\{}\frac{2mk\lambda(\kappa_{i}-\kappa_{f})}
{\Pho+\mtri}(1+\frac{\Pho}{\mtri})\Big{\}}+\frac{F_{M}^{H}}{\Pho+
\mtri}\Big{\{}{\lambda}\frac{\nu-m}{4\mtri}(\frac{2m(2+\kappa_{i})}
{\mtri}+\frac{4mk}{\zeta})+
{\lambda\kappa_{f}}\frac{mk}{\zeta}\frac{\Pho(m-\nu)}{\mtri^{2}}\Big{\}}+
\nonumber\\
&&F_{M}^{L}\Big{\{}\frac{1-\lambda}{4(\Pho+\mtri)}(-2+
\frac{2m}{\mtri}(1-\frac{\Pho+\nu+k}{m}))\Big{\}}+
\nonumber\\
&&\frac{F_{A}^{H}}{\Pho+\mtri}\Big{\{}(\frac{2m}{\mtri}-\frac{2mk}
{\zeta})-{\kappa_{f}}\frac{2mk}{\zeta}(1+\frac{\Pho}{\mtri})\Big{\}}+
F_{A}^{\prime}\Big{\{}-4m\Big{\}}
\nonumber\\
f_{9}&=&\frac{F_{V}^{H}}{\Pho+\mtri}\Big{\{}-\lambda(2+\kappa_{i})
\frac{m}{\mtri}-\frac{2mk}{\zeta}(\lambda+2\dprod{\hat{\nu}}{\hat{k}})-
\frac{4m\nu}{\zeta}-\kappa_{f}\frac{2mk}{\zeta}(\lambda+
\frac{\dprod{\hat{\nu}}{\vec{k}}+\nu}{\mtri})\Big{\}}+
F_{V}^{L}\Big{\{}\frac{1-\lambda}{\Pho+\mtri}\Big{\}}+
\nonumber\\
&&F_{V}^{\prime}\Big{\{}-4m\frac{\nu+\dprod{\hat{\nu}}{\vec{k}}}
{\Pho+\mtri}\Big{\}}+\frac{F_{M}^{H}}{\Pho+\mtri}\Big{\{}-(1+
\lambda)\frac{m\nu}{\mtri^{2}}+\lambda\frac{m^{2}}{\mtri^{2}}+
\frac{m\nu}{\zeta}(\frac{k}{\mtri}-4)+\frac{\lambda
m k}{\zeta}\frac{m-\nu-k}{\mtri}+ \nonumber\\
&&\frac{\dprod{\vec{\nu}}{\vec{k}}}{\zeta}(-\frac{2m}{\mtri}(1-
\frac{m}{\nu}))-\frac{m\dprod{\hat{\nu}}{\vec{k}}}{\mtri^{2}}+
\lambda\kappa_{i}\frac{m(m-\nu)}{2\mtri^{2}}+\kappa_{f}
\frac{mk}{\zeta}\frac{m\nu-\nu^{2}+\dprod{\vec{\nu}}{\hat{k}}(m-k)}{
\mtri^{2}}+ \nonumber\\
&&\lambda\kappa_{f}\frac{mk}{\zeta}(\frac{m-\nu-k}{\mtri}-\frac{s^{2}-
2\nu^{2}-2\dprod{\vec{\nu}}{\vec{k}}+
2\Pho k}{2\mtri^{2}})\Big{\}}+\frac{F_{M}^{L}(1-\lambda)}{\Pho+
\mtri}\Big{\{}\frac{1}{2}(1+\frac{\Pho}{\mtri})-\frac{m}{2\mtri}+
\frac{k+\nu}{2\mtri}\Big{\}}+
\nonumber\\
&&\frac{F_{M}^{\prime}}{\Pho+\mtri}\Big{\{}-2m\nu(1-\frac{m-\nu-
\dprod{\hat{\nu}}{\vec{k}}-\Pho}{\mtri})+\frac{2m^{2}
\dprod{\hat{\nu}}{\vec{k}}}{\mtri}
\Big{\}}+\frac{F_{A}^{H}}{\Pho+\mtri}\Big{\{}-\frac{2m}{\mtri}(1-
\frac{\mtri
k}{\zeta})-\kappa_{f}\lambda\frac{2mk}{\zeta}(\frac{\nu+
\dprod{\hat{\nu}}{\vec{k}}}{\mtri})
\nonumber\\
&&+\frac{4m(\Pho+\mtri)}{\zeta}\Big{\}}+F_{A}^{\prime}\Big{\{}4m\Big{\}}
\nonumber\\
f_{10}&=&\frac{F_{V}^{H}}{\Pho+\mtri}\Big{\{}-\frac{4\kappa_{f}\lambda
m^{2}}{\zeta}\frac{k}{\mtri}\Big{\}}+\frac{F_{M}^{H}}{\Pho+\mtri}
\Big{\{}\frac{2m^{2}}{\mtri^{2}}+\lambda\frac{4m^{2}}{\zeta}
\frac{k}{\mtri}(-1+\kappa_{f}\frac{3\Pho}{4\mtri}-
\frac{\kappa_{f}}{4})\Big{\}}+
\nonumber\\
&&\frac{F_{A}^{H}}{\Pho+\mtri}\Big{\{}-\frac{4m^{2}}{\zeta}(2+
\kappa_{f}\frac{k}{\mtri})\Big{\}}+\frac{F_{A}^{\prime}}{\Pho+
\mtri}\Big{\{}-8m^{2}+\frac{16\mtri
m^{3}(1+\varepsilon)}{m_{\pi}^{2}-\Qlepml^{2}}\Big{\}}+ \nonumber\\
&&\frac{F_{P}^{H}}{\Pho+\mtri}\Big{\{}\frac{8m^{2}}{\zeta}+
\frac{8m^{2}}{m_{\pi}^{2}-\Qlepml^{2}}+\kappa_{f}\frac{4m^{2}}
{\zeta}\frac{k}{\mtri}\Big{\}}
\nonumber\\
f_{11}&=&\frac{F_{V}^{\prime}}{\Pho+\mtri}\Big{\{}8m^{2}\Big{\}}+
\frac{F_{M}^{H}}{\Pho+\mtri}\Big{\{}\frac{2m^{2}}{\mtri^{2}}\Big{\}}
\nonumber\\
f_{12}&=&\frac{F_{V}^{H}}{\Pho+\mtri}\Big{\{}\frac{8m^{2}}{\zeta}
(\lambda\frac{k}{\nu}-1)+\kappa_{f}\frac{4m^{2}}{\zeta}\frac{k}
{\mtri}(\lambda\frac{k}{\nu}-1)\Big{\}}+\frac{F_{V}^{\prime}}{\Pho+
\mtri}\Big{\{}8m^{2}(\lambda\frac{k}{\nu}-1)\Big{\}}+
\nonumber\\
&&\frac{F_{M}^{H}}{\Pho+\mtri}\Big{\{}\frac{2m^{2}}{\mtri^{2}}
(\lambda\frac{k}{\nu}-1)-\frac{8m^{2}}{\zeta}+\frac{4\lambda
m^{2}}{\zeta}\frac{k}{\mtri}(1-\frac{m}{\nu})-\kappa_{f}\frac{4m^{2}}
{\zeta}\frac{k}{\mtri}+
\nonumber\\
&&\lambda\kappa_{f}\frac{2m^{2}}{\zeta}\frac{k}{\mtri}(-\frac{1}{2}+
\frac{k}{\mtri}(1-\frac{m}{\nu})+\frac{2\nu-\Pho}{2\mtri})\Big{\}}+
\frac{F_{M}^{\prime}}{\Pho+\mtri}\Big{\{}-4m^{2}(1-\frac{m-\nu-\Pho}
{\mtri})+4\lambda
m^{2}\frac{k}{\mtri}(1-\frac{m}{\nu})\Big{\}}+ \nonumber\\
&&\frac{F_{A}^{H}}{\Pho+\mtri}\Big{\{}\kappa_{f}\frac{4m^{2}}{\zeta}
\frac{k}{\mtri}(-\lambda+\frac{k}{\nu})\Big{\}}
\nonumber\\
f_{13}&=&\frac{F_{V}^{H}}{\Pho+\mtri}\Big{\{}\frac{4mk}{\zeta}(1+
\kappa_{f}\frac{k}{\mtri})(\lambda+\dprod{\hat{\nu}}{\hat{k}})\Big{\}}+
\frac{F_{V}^{\prime}}{\Pho+\mtri}\Big{\{}4mk(\lambda+\dprod{\hat{\nu}}
{\hat{k}})\Big{\}}+\frac{F_{M}^{H}}{\Pho+\mtri}\Big{\{}\frac{m\dprod{
\hat{\nu}}{\vec{k}}}{\mtri^{2}}+
\nonumber\\
&&\frac{2mk}{\zeta}(\frac{2\dprod{\hat{\nu}}{\hat{k}}(\nu-m)-k}{2\mtri})+
\frac{\lambda
mk}{\mtri^{2}}+\frac{2\lambda mk}{\zeta}(\frac{2(\nu-m)+k}{2\mtri})+
\nonumber\\ &&\kappa_{f}\frac{2mk}{\zeta}(\frac{2((\lambda-1)(\nu
k-\dprod{\vec{\nu}}{\vec{k}})-2\Pho(\nu-\lambda
k)-2m(\dprod{\hat{\nu}}{\vec{k}}+\lambda k))+\lambda
s^{2}}{4\mtri^{2}}+\frac{\lambda k-\nu}{2\mtri}){\Big{\}}}+ \nonumber\\
&&\frac{F_{M}^{\prime}}{\Pho+\mtri}\Big{\{}\frac{2m}{\mtri}(
\dprod{\hat{\nu}}{\hat{k}}(\nu
k-mk)-\lambda k(m-\nu))\Big{\}}+ \nonumber\\
&&\frac{F_{A}^{H}}{\Pho+\mtri}\Big{\{}\kappa_{f}\frac{2mk}{\zeta}
\frac{k}{\mtri}(1+\lambda\dprod{\hat{\nu}}{\hat{k}})\Big{\}}+
F_{A}^{\prime}\Big{\{}-4m\Big{\}}
\nonumber\\
f_{14}&=&\frac{F_{V}^{\prime}}{\Pho+\mtri}\Big{\{}8m^{2}+
(\kappa_{i}-\kappa_{f})\frac{4m^{2}k}{\mtri}\Big{\}}
\nonumber\\
f_{15}&=&\frac{F_{V}^{H}}{\Pho+\mtri}\Big{\{}\frac{16m^{3}}{\zeta\nu}(1+
\kappa_{f}\frac{k}{2\mtri})\Big{\}}+F_{V}^{\prime}\Big{\{}
\frac{16m^{3}}{(\Pho+\mtri)\nu}\Big{\}}+\frac{F_{M}^{H}}{\Pho+
\mtri}\Big{\{}\frac{4m^{3}}{\mtri^{2}\nu}(1-(2+\kappa_{f}\frac{k}
{\mtri})\frac{\mtri
m}{\zeta})\Big{\}}+ \nonumber\\
&&F_{M}^{\prime}\Big{\{}-\frac{8m^{4}}{\mtri(\Pho+\mtri)\nu}\Big{\}}+
F_{A}^{H}\Big{\{}\kappa_{f}\frac{8m^{3}\lambda
k}{\mtri(\Pho+\mtri)\zeta\nu}\Big{\}}
\end{eqnarray}

\section{Wave Function characteristics}\label{bind_en}

Table (\ref{wavetable}) lists several important quantities that depend
on the potential used to generate the wave function.  They are
respectively the binding energy given by the wave functions, various
partial wave probabilities and the numerical normalization
$\ip{\psi}{\psi}_{\mathrm{num}}$ of the wave functions.  The numerical
normalizations of all the wave functions are not unity because the
antisymmetrization of the wave functions is projected on a finite set
of states.  The experimental binding energy $E_{b}$ of
$\mathrm{^{3}He}$ is 7.72 MeV and $\mathrm{^{3}H}$ is 8.48 MeV
\cite{Wapstra85}.

\begin{table}[hbt]
\begin{tabular}[c]{|c|c|c|c|c|c|c|}
\hline
Potential & $E_{b}$ & $P(S)$ & $P(S^{\prime})$ & $P(P)$ & $P(D)$ &
$\ip{\psi}{\psi}_{\mathrm{num}}$ \\ \hline
Bonn A & 8.29 & $92.59\%$& $1.23\%$ &$0.030\%$&$6.14\%$& 0.994 \\
\hline
Bonn B & 8.10 & $91.61\%$& $1.19\%$ &$0.044\%$&$7.16\%$& 0.993 \\
\hline
CD Bonn $(\mathrm{^{3}He})$ & 7.91 & $91.61\%$& $1.35\%$
&$0.041\%$&$7.01\%$& 0.993 \\ \hline
CD Bonn $(\mathrm{^{3}H})$ & 7.96 & $91.63\%$& $1.31\%$
&$0.041\%$&$7.01\%$& 0.993 \\ \hline
Nijmegen I & 7.66 & $90.31\%$& $1.29\%$ &$0.065\%$&$8.34\%$& 0.990 \\
\hline
Paris & 7.38 & $90.11\%$& $1.40\%$ &$0.069\%$&$8.42\%$& 0.988 \\
\hline
AV14 & 7.58 & $89.86\%$& $1.15\%$ &$0.082\%$&$8.90\%$& 0.987 \\ \hline
\end{tabular}
\vspace{0.5cm}
\caption{\label{wavetable}
Some important quantities of trinucleon wave functions.  The binding
energy $E_{b}$ is in MeV.  $P(S)$ denotes the probability of S-wave
component of the wave function and so on.}
\end{table}


\begin{thebibliography}{10}
\bibitem{E592}
D. H. Wright, TRIUMF research proposal E592, (1990).
\bibitem{Wright99}
D. H. Wright et al, Few Body Systems Suppl. {\bf 12}, 275  (2000).
\bibitem{Mukhopadhyay77}
N. C. Mukhopadhyay, Physics Reports {\bf 30}, 1 (1977).
\bibitem{Measday01}
D. F. Measday, Physics Reports {\bf 354}, 243 (2001).
\bibitem{Klieb81}
L. Klieb and H. P. C. Rood, Nuclear Physics {\bf A356}, 483 (1981).
\bibitem{Kliebthesis}
L. Klieb, Ph. D. thesis, University of Groningen, The Netherlands, (1982).
\bibitem{Congletonthesis}
J. G. Congleton, Ph. D. thesis, University of British Columbia, (1992), 
unpublished.
\bibitem{Congleton93}
J. G. Congleton and H. W. Fearing, Nuclear Physics {\bf A552}, 534 (1993).
\bibitem{Congleton96}
J. G. Congleton and E. Truhlik, Phys. Rev. C {\bf 53}, 956 (1996).
\bibitem{Govaerts00}
J. Govaerts and J-L. Lucio-Martinez, Nuclear Physics {\bf A678}, 110 (2000).
\bibitem{Adler66}
S. L. Adler and Y. Dothan, Phys. Rev. {\bf 151}, 1267 (1966).
\bibitem{Rood65}
H. P. C. Rood and H. A. Tolhoek, Nuclear Physics {\bf 70}, 658 (1965).
\bibitem{Christillin80}
P. Christillin and S. Servadio, Nuovo Cimento {\bf 42A}, 165 (1977).
\bibitem{Primakoff79}
H. Primakoff, Nuclear Physics {\bf A317}, 219 (1979).
\bibitem{Kim65}
C. W. Kim and H. Primakoff, Phys. Rev. {\bf 140}, B566 (1965).
\bibitem{Kim65s}
C. W. Kim and H. Primakoff, Phys. Rev. {\bf 139}, B1447 (1965).
\bibitem{Fearing80}
H. W. Fearing, Phys. Rev. C {\bf 21}, 1951 (1980).
\bibitem{Klieb85}
L. Klieb, Nucl. Phys. {\bf A442}, 721 (1985).
\bibitem{Groom00}
D. E. Groom et al. (Particle Data Group). Eur. Phys. Jour. 
{\bf C15}, 1 (2000).
\bibitem{Schadow00}
W. Schadow, W. Sandhas, J. Haidenbauer and A. Nogga, 
Few Body Systems {\bf 28}, 241 (2000).
\bibitem{Schadowthesis}
W. Schadow, Ph. D. thesis, Bonn University, Germany, BONN-IR-97-17, (1997).
\bibitem{Glockle}
W. Gl\"{o}ckle, {\it The Quantum Mechanical Few Body Problem}, 
(Springer-Verlag, 1983).
\bibitem{Arfken}
G. B. Arfken and H. J. Weber, {\it Mathematical Methods for 
Physicists} (Academic Press, 1995).
\bibitem{Brink}
D. M. Brink and G. R. Satchler, {\it Angular Momentum} (Oxford 
Science Publications, 1993).
\bibitem{Delorme}
J. Delorme, {\it Mesons in Nuclei}, edited by M. Rho and D. H. 
Wilkinson (North-Holland, Amsterdam, 1979), Vol I, 107.
\bibitem{Peterson67}
E. A. Peterson, Phys. Rev. {\bf 167}, 971 (1968).
\bibitem{Lahiff97}
A. D. Lahiff and I. R. Afnan, Phys. Rev. C {\bf 56}, 2387 (1997).
\bibitem{Ackerbauer98}
P. Ackerbauer et al, Physics Letters {\bf B417}, 224 (1998).
\bibitem{Kameyama89}
H. Kameyama, M. Kamimura and Y. Fukushima, Phys. Rev. C {\bf 40}, 974 (1989).
\bibitem{Fearing75}
H. W. Fearing, Phys. Rev. Lett. {\bf 35}, 79 (1975).
\bibitem{Wapstra85}
A. H. Wapstra and G. Audi, Nuclear Physics {\bf A595}, 409 (1995).

\end{thebibliography}
\end{document}